%% file: main.tex
\documentclass[sigconf]{acmart}

\usepackage{multirow}
\usepackage{xcolor}
\usepackage{longtable}
\usepackage{array}
\usepackage{float}
\usepackage{colortbl}
\usepackage{enumitem}
\usepackage{fontawesome5}
\usepackage{CJK}
\usepackage{framed}
\usepackage{makecell}  
\usepackage{booktabs}  
\usepackage{fancybox}
\usepackage{acmart-taps}

\newlength\MAX  
\newcommand*\Chart[1]{\rlap{\textcolor{black!20}{\rule{\MAX}{2ex}}}\rule{#1\MAX}{2ex}}

\AtBeginDocument{%
  }

\copyrightyear{2025}
\acmYear{2025}
\setcopyright{none}
\acmConference[CHI '25]{CHI Conference on Human Factors in Computing Systems}{April 26-May 1, 2025}{Yokohama, Japan}
\acmBooktitle{CHI Conference on Human Factors in Computing Systems (CHI '25), April 26-May 1, 2025, Yokohama, Japan}\acmDOI{10.1145/3706598.3713567}
\acmISBN{979-8-4007-1394-1/25/04}

\begin{document}
\begin{CJK}{UTF8}{gbsn}

\title[Human-Precision Medicine Interaction: Public Perceptions of Polygenic Risk Score]{Human-Precision Medicine Interaction: Public Perceptions of Polygenic Risk Score for Genetic Health Prediction}

\author{Yuhao Sun}
\authornote{The author is known as 孙宇浩.}

\email{yuhao.sun@ed.ac.uk}
\orcid{0000-0002-4053-6032}
\affiliation{%
  \institution{University of Edinburgh}
  \city{Edinburgh}
  \state{Scotland}
  \country{United Kingdom}
}

\author{Albert Tenesa}
\email{albert.tenesa@ed.ac.uk}
\orcid{0000-0003-4884-4475}
\affiliation{%
  \institution{University of Edinburgh}
  \city{Edinburgh}
  \state{Scotland}
  \country{United Kingdom}
}

\author{John Vines}
\email{john.vines@ed.ac.uk}
\orcid{0000-0003-4051-3356}
\affiliation{%
  \institution{University of Edinburgh}
  \city{Edinburgh}
  \state{Scotland}
  \country{United Kingdom}
}

\begin{abstract}
Precision Medicine (PM) transforms the traditional ``one-drug-fits-all'' paradigm by customising treatments based on individual characteristics, and is an emerging topic for HCI research on digital health. A key element of PM, the Polygenic Risk Score (PRS), uses genetic data to predict an individual's disease risk. Despite its potential, PRS faces barriers to adoption, such as data inclusivity, psychological impact, and public trust. We conducted a mixed-methods study to explore how people perceive PRS, formed of surveys (n=254) and interviews (n=11) with UK-based participants. The interviews were supplemented by interactive storyboards with the ContraVision technique to provoke deeper reflection and discussion. We identified ten key barriers and five themes to PRS adoption and proposed design implications for a responsible PRS framework. To address the complexities of PRS and enhance broader PM practices, we introduce the term Human-Precision Medicine Interaction (HPMI), which integrates, adapts, and extends HCI approaches to better meet these challenges.
\end{abstract}




\begin{teaserfigure}
  \includegraphics[width=\textwidth]{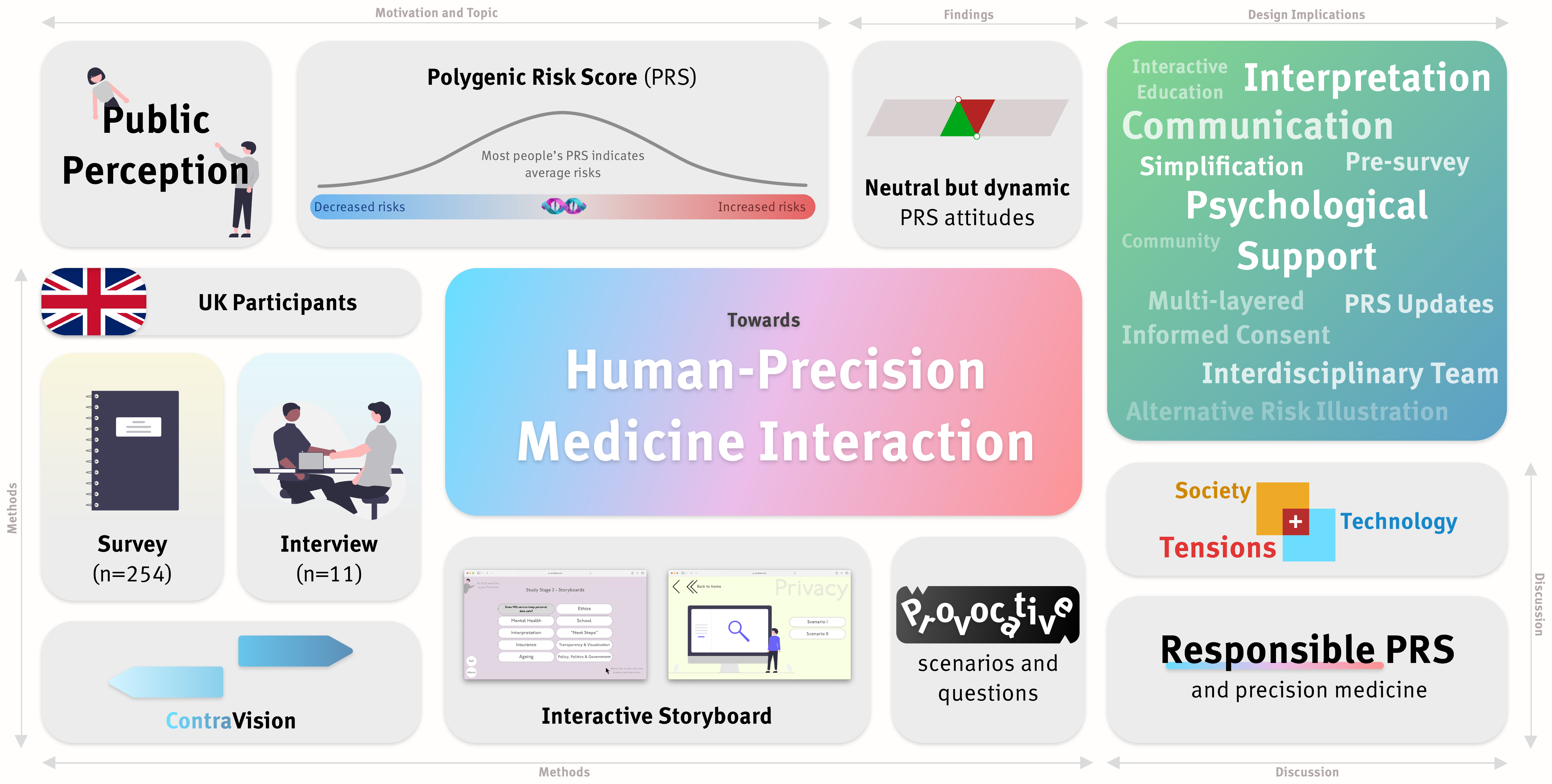}
  \caption{\textbf{Teaser Bento: A Snapshot of Our Paper.}}
  \Description{The figure is arranged in a "Bento Box" style layout to provide a structured overview of the key points from our paper. In the top left compartment, it presents the "Motivation and Topic" (Public Perception of Polygenic Risk Score). The top right compartment focuses on "Findings" (neutral but dynamic PRS attitudes) and "Design Implications." In the bottom left compartment, the "Methods" section (UK participants, survey, interview, ContraVision technique, interactive storyboard, and provocative scenarios and questions) is detailed. The bottom right compartment is dedicated to "Discussion" (society and technology tensions, responsible PRS and precision medicine). At the centre of the Bento Box, the core concept we proposed of "Human-Precision Medicine Interaction" is prominently featured.}
  \label{fig:teaser}
\end{teaserfigure}

\maketitle

\section{Introduction}
\begin{quote}
    ``\textit{To enable a new era of medicine through research, technology, and policies that empower patients, researchers, and providers to work together toward development of individualized care.}''  \flushright-- Mission Statement, Precision Medicine Initiative, 2015 \cite{obamaPM:online}
\end{quote}
Precision Medicine (PM) is an innovative approach to healthcare that customises and personalises medical treatment to the characteristics of each individual \cite{ashley2016towards,ginsburg2018precision,wang2023precision,johnson2021precision}. By utilising detailed health information, PM aims to optimise prevention, diagnosis, and treatment strategies to enhance healthcare outcomes -- and revolutionise the ``one-drug-fits-all'' situation. Polygenic Risk Score (PRS), as a representative of the PM revolution, is a technology that estimates an individual's genetic risk for various diseases before their onsets \cite{lewis2017prospects,lewis2020polygenic,cross2022polygenic}. Researchers identify common genetic variants associated with specific diseases and assign them weights based on their impact on disease risk \cite{choi2019prsice}. These weighted effects are summed to provide a comprehensive risk score, indicating the likelihood of developing certain diseases based on one's genetic makeup \cite{choi2019prsice,choi2020tutorial}.

Technically, PM applications, such as PRS, heavily rely on the ability to effectively process, interpret, and communicate the insights derived from complex, multi-dimensional healthcare datasets \cite{johnson2021precision,simpkin2019communicating} -- mostly biological data with genetics in a central position \cite{delpierre2023precision,xu2021exploring}. This task involves significant computational challenges through machine learning, high-performance computing, and other various specialised AI methodologies \cite{bhinder2021artificial,filipp2019opportunities}. On the other hand, there are still open questions as to how these advanced PM technologies can be integrated into real-world healthcare practices, ensuring that information is used and presented in a way that is understandable, usable, and responsibly empowering for both patients and healthcare professionals \cite{xue2016precision,xu2021exploring,kohane2015ten,kerr2018limits} -- here, HCI could play a crucial role \cite{rundo2020recent}.

PRS provides an exemplary case for exploring the intersection of HCI and PM. As a technically advanced and well-established application within the PM landscape, PRS serves as a comprehensive and thought-provoking example for examining both the potential and the challenges of integrating genetic data into PM. Despite its potential to revolutionise PM by enabling early disease risk detection and tailored prevention strategies, PRS faces several challenges and barriers to widespread adoption \cite{peck2022people,martin2019clinical,kachuri2024principles} and responsible use \cite{polygenic2021responsible}. These include concerns about the accuracy of PRS for individuals from diverse genetic backgrounds due to the predominant reliance on European-centric genotype data \cite{martin2019clinical,kachuri2024principles}, and possible negative emotional reactions to receiving risk information \cite{peck2022people}. The potential use of PRS under different contexts, such as embryo selection \cite{turley2021problems},  insurance \cite{yanes2024future} or school settings (academic achievements \cite{selzam2017genome} and bullying \cite{schoeler2019multi}), invites important discussions on privacy, fairness, and the ethical implications of such technologies. In the UK, for example, the \textit{care.data} programme scandal has profoundly shaken public trust in healthcare data sharing \cite{sterckx2016you}. The public, thus, has adopted a more cautious and often sceptical stance towards such initiatives \cite{lounsbury2021opening,kerasidou2023data}. However, trust remains a fundamental prerequisite for the acceptance and success of data-driven healthcare technologies \cite{gille2020public,gille2021public,horn2020sharing}. Furthermore, consumer-facing PRS services, including those provided by Direct-To-Consumer Genetic Testing (DTC-GT) companies like 23andMe and Ancestry \cite{su2013direct,park2023polygenic}, add another layer of complexity regarding privacy and the perceived reliability of results \cite{majumder2021direct}. 

Our research was motivated by these concerns, and set out to understand what the perceived barriers, benefits and challenges potential users and beneficiaries see in relation to adopting and using PRS-based services in their everyday lives. We also set out to explore how participants' individual experiences, in particular their experiences of sharing personal data and their use of health services, might affect their decision-making around potentially adopting PRS. We conducted a mixed-methods study involving a 254-person survey and 11 one-hour one-to-one interviews which were scaffolded by interactive storyboards developed following the ContraVision technique \cite{mancini2010contravision}. The survey aimed to understand public perceptions of PRS and explore potential barriers to its use, providing quantitative and qualitative data. The interviews provided nuanced qualitative insights into how people's experiences and stories affect their informed decisions regarding PRS. Our survey findings identified ten key barriers to PRS adoption and uncovered a complex landscape of opinions on the application of PRS in socially sensitive contexts. While participants generally maintained a neutral stance on PRS, specific areas like embryo selection and sperm/egg donation attracted notable interest, in contrast to negative opinions on its use in education, insurance, and dating due to societal and ethical concerns. Then, our interview findings demonstrated key challenges, including the complexity of interpreting PRS results, demographic bias in genetic databases, and concerns about psychological preparedness. Participants highlighted the need for clear information and support, particularly regarding the emotional and ethical aspects, while expressing varying degrees of scepticism about trust in the healthcare system and data privacy.

Our paper contributes to the field of HCI by reporting the above empirical findings and developing design recommendations and implications for the future PRS and PM more broadly. We also discuss how a responsible PRS framework can address systemic complexities by prioritising data diversity, trust, and ethical governance. These highlight a crucial yet underexplored need -- adapting and extending HCI perspectives to manage the unique complexities of PM, such as its reliance on highly individualised data and the dynamic nature of patient care. From this, we outline potential future directions by proposing Human-Precision Medicine Interaction (HPMI) as a potential new area of enquiry within HCI that emphasises public engagement in co-developing and implementing PM technologies.

\section{Related Work}

\subsection{Precision Medicine (PM) and Human-Computer Interaction (HCI)}

Precision Medicine (PM)\footnote{We use the term \textit{Precision Medicine (PM)} to refer to all approaches or terminologies that tailor medical decisions, practices, and interventions based on a patient's individual characteristics.} tailors treatments based on an individual's health-related data (such as genetic, environmental, and lifestyle factors), using data analysis to improve diagnosis and treatment efficacy \cite{ashley2016towards,ginsburg2018precision,wang2023precision,johnson2021precision}. The core of PM lies in the collection, analysis and interpretation of data. Different from a ``one-drug-fits-all'' strategy, PM leverages the analysis of genetic and other health information and other biomarkers to provide ``the right drug at the right dose to the right patient'' \cite{collins2015new}. For instance, in oncology, genetic analysis of tumours enables more targeted therapies \cite{malone2020molecular}. Another example of PM is using the weaknesses in cancer cells' DNA repair process to make treatment more effective for some breast cancer patients \cite{robson2017olaparib}.  With advancements in technology and the accumulation of data, PM is expected to promote the personalisation and precision of healthcare further, offering better health management and treatment options for patients. Although PM as a field was established around 1998 with the approval of the first application in the breast cancer field \cite{de2021translational}, it has been gaining significant momentum with the U.S. Precision Medicine Initiative since 2015 \cite{collins2015new}.

PM is inherently interdisciplinary, involving more than a hundred disciplines \cite{xu2021exploring} such as genomics, data science and machine learning, and clinical research \cite{seyhan2022current}. In recent years, the CHI community has seen a burgeoning interest in PM. Although there are only a few papers that explicitly use the term ``precision medicine'' or its alternatives\footnote{The alternatives of PM include but are not limited to ``personalised medicine,'' ``customised medicine,'' or ``individualised medicine.'' These terms may be used differently depending on the context. Generally, they share similar meanings with PM.} to define their work, the field has implicitly explored relevant themes through various HCI studies. For example, Calisto et al. focus on tailoring communication between intelligent systems and clinicians in breast cancer diagnosis, significantly reducing medical errors and enhancing decision-making \cite{calisto2023assertiveness}. Similarly, Mitchell et al. integrate machine learning with expert systems to provide tailored nutrition recommendations for individuals with type 2 diabetes, illustrating how HCI research can make complex data-driven insights actionable for patients \cite{g2021reflection}. However, these studies also raise important considerations about the balance between automation and human oversight in healthcare. It must carefully navigate the ethical and practical challenges of integrating these systems into everyday clinical practice, ensuring that the technology enhances rather than overshadows the human elements of care. For example, again in oncology, Verma et al. discuss how physicians' trust in AI is more grounded in their contestable experiences with AI in actual practice rather than general acceptance \cite{verma2023rethinking}.

Beyond explicit PM research, many contributions in this field focus on adjacent issues critical to PM. A central theme is \textbf{data privacy and security}, by designing interfaces that enhance privacy protections and provide fine-grained control over data consent, thereby supporting users to effectively manage their sensitive health information \cite{baig2020m,lee2024priviaware,malki2024exploring}. Additionally, researchers have evaluated privacy risks and integrated regulatory frameworks such as GDPR into the design of tracking technologies, ensuring the safeguarding of user data and fostering trust in digital health applications \cite{mehrnezhad2021caring,tazi2024we}. Another key area is \textbf{patient engagement and empowerment}, where personalised interfaces have been designed to help patients better understand and manage their health data, thereby enhancing their proactive involvement in medical decision-making \cite{hao2024advancing,haldar2019beyond,yoo2024patient}. Moreover, research in this area has facilitated effective communication between patients and healthcare teams, supporting active patient participation in customised treatment plans and collaborative decision-making \cite{liaqat2024promoting,mishra2016not,bedmutha2024conversense}.

\subsection{Polygenic Risk Scores (PRS)}

\begin{figure*}
      \includegraphics[width=\textwidth]{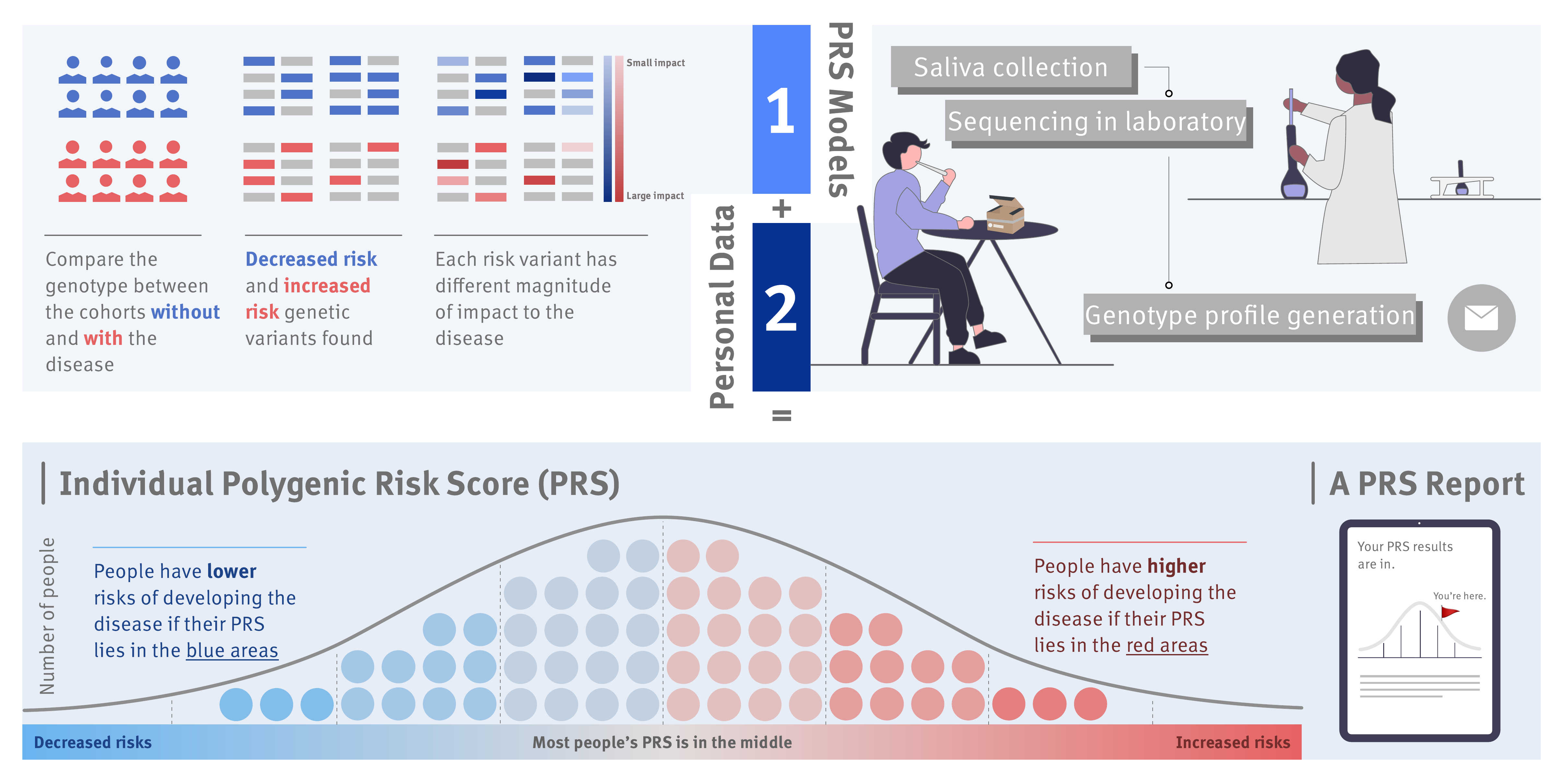}
  \caption{\textbf{Calculation and Interpretation of Polygenic Risk Score (PRS).} The calculation of a PRS involves two main steps. First, genetic variants associated with increased or decreased risks are identified by comparing genotype data from cohorts with and without a specific disease, along with the magnitude of their impact. Second, individuals generate their personal genotype profiles through Direct-to-Consumer Genetic Testing (DTC-GT) services, which require saliva collection and DNA sequencing. PRS typically follow a normal distribution, with most individuals falling around the average risk and some at the tails, indicating either decreased or increased risks. Sometimes, PRS reports provide supplementary information to help individuals interpret the results.
  % \cite{Polygeni94:online}
  }
  \Description{The figure illustrates the process of calculating and interpreting a Polygenic Risk Score (PRS). The process is divided into two primary steps. The first step is the identification of genetic variants. It involves identifying genetic variants associated with increased or decreased risks for a specific disease. This is done by comparing genotype data from two groups - one with the disease and one without it - alongside measuring the magnitude of each variant's impact. The second step is the generation of personal genotype profiles. Individuals obtain their genotype profiles through Direct-to-Consumer Genetic Testing (DTC-GT) services, which involve collecting a saliva sample and conducting DNA sequencing to analyse the genetic information. The figure further indicates that PRS values typically exhibit a normal distribution pattern. Most individuals have a PRS near the average, which suggests an average risk, while some individuals fall into the lower or upper tails of the distribution, indicating a decreased or increased risk, respectively. The figure also highlights that PRS reports often include additional information to aid in interpreting these scores.}
  \label{fig:prs-method}
\end{figure*}

Polygenic Risk Score (PRS)\footnote{We use the term \textit{Polygenic Risk Score (PRS)} to refer to all technologies and methods that aggregate genetic data to provide individual risk estimates.}, a genetics-based health prediction technology, is representative of PM. PRS predicts an individual's genetic risk for diseases before the onset of the disease \cite{lewis2017prospects,lewis2020polygenic,cross2022polygenic}. Figure \ref{fig:prs-method} illustrates how a PRS is generated, which starts by constructing a PRS model for the specific disease. Researchers first conduct Genome-Wide Association Studies (GWAS), which identify genetic variants linked to specific diseases \cite{choi2020tutorial,kullo2022polygenic}. Each identified variant is given a weight based on its impact on disease risk \cite{choi2019prsice}. PRS is calculated by combining an individual's weighted genetic variants, reflecting their likelihood of developing certain diseases \cite{choi2019prsice,choi2020tutorial}. Then, the individual genotype profile is needed to generate the PRS. The typical process often starts with Direct-To-Consumer Genetic Testing (DTC-GT) services like 23andMe\footnote{\href{https://www.23andme.com/en-gb/}{https://www.23andme.com/en-gb/}} or Ancestry\footnote{\href{https://www.ancestry.co.uk/}{https://www.ancestry.co.uk/}}, which offer genetic testing through saliva samples for a fee \cite{ormondroyd2022genomic}. After receiving their genotype data from these services, individuals can choose to share their genetic data with third-party providers for PRS analysis \cite{folkersen2020impute}. The analysis process varies depending on the provider, and generally involves steps such as data upload, genetic imputation, risk calculation, and report generation, with options for privacy settings and consultation. Some DTC-GT services also include PRS analysis as part of their health reports \cite{nolan2023direct}. Additionally, individuals with sufficient expertise in biostatistics and genetics can use specialised PRS analysis software through academic or research channels \cite{choi2019prsice,euesden2015prsice}. PRS is typically presented as percentiles, following a normal distribution \cite{wray2007prediction,dudbridge2013power}. Most individuals find their scores in the middle of the distribution, indicating an average risk of disease compared to the overall database \cite{lambert2019towards}. Some might find their scores at the two ends, indicating either a lower (decreased) or higher (increased) risk, depending on their specific position \cite{lambert2019towards}. However, this distribution can vary depending on the specific population and traits being analysed \cite{chatterjee2016developing}.

PRS has the potential to transform PM by facilitating early detection of disease risks and enabling customised prevention strategies. There are currently various PRS models for diseases such as coronary artery disease, type 1 and 2 diabetes, obesity, prostate cancer, breast cancer, and Alzheimer's disease, with further clinical evaluation necessary \cite{patel2023multi,ho2020european,lambert2019towards,chatterjee2016developing}. By providing insight into genetic risk, PRS can encourage lifestyle changes, regular monitoring, and preventive measures to reduce disease risk \cite{khera2016genetic,torkamani2018personal}. Additionally, PRS allows for personalised health management plans tailored to an individual's genetic profile, making healthcare more efficient and targeted \cite{torkamani2018personal,chatterjee2016developing}. With knowledge of their genetic risk, individuals can make better-informed decisions about diet, exercise, medication, and health screenings \cite{lewis2021polygenic}. PRS also promotes greater health awareness, encouraging proactive health behaviours and enhancing overall well-being \cite{chatterjee2016developing}. Moreover, it offers valuable insights for family health planning, enabling relatives to collaboratively develop strategies, particularly for managing hereditary conditions \cite{khera2016genetic,lewis2021polygenic}.

While DTC-GT services generally provide consumers with relatively clear-cut genetic information -- ancestry breakdowns or single-gene trait analysis, PRS explores further into the \textbf{predictive} domain. This shift from descriptive to predictive data introduces significant complexity, raising critical concerns that challenge both its adoption and responsible use \cite{peck2022people,martin2019clinical,kachuri2024principles,polygenic2021responsible}. The predictive nature of PRS brings inherent uncertainty and variability. This is particularly problematic since PRS relies on existing genotype databases, which are predominantly European-centric, reducing its accuracy for individuals from diverse genetic backgrounds and exacerbating health disparities \cite{martin2019clinical,kachuri2024principles}. Moreover, the psychological burden of receiving risk predictions can cause anxiety and influence decisions \cite{peck2022people}. The ethical landscape becomes even more intricate when considering the application of PRS in sensitive areas such as embryo selection \cite{turley2021problems}, insurance \cite{yanes2024future}, and education \cite{selzam2017genome,schoeler2019multi}. In these contexts, PRS not only raises questions about privacy and fairness but also about the potential for misuse and discrimination. The consumer-driven nature of many PRS services, especially those offered by DTC-GT services, introduces additional complexity \cite{su2013direct,park2023polygenic,majumder2021direct}. Individuals interpreting complex risk information alone increases the risk of misuse \cite{nolan2023direct,horton2019direct} -- with only a few consumers seeking genetic counselling services \cite{koeller2017utilization}. Additionally, concerns about data privacy, risk assessment reliability, and lack of regulatory oversight add to these complexities.

\subsection{Public Perceptions of Genetic Testing and Data}

Given the limited literature on PRS attitudes, we focused on the broader research on \textbf{genetic testing and data}. Zhang et al. identified two main categories of public perceptions of genetic information research \cite{zhang2021public}. The first category focuses on genetic testing, primarily within clinical care and reproductive services, with some studies distinguishing between medical and non-medical outcomes. These studies suggest that support is lower for traits central to identity (e.g., height or talents) and higher for traits viewed as disease-related. The second category investigates how consent structures, incentives, and privacy concerns influence the willingness to provide genetic information for research. Respondents in the studies under this category were concerned about discrimination by employers and insurers, and the potential access to genetic data by commercial entities and government agencies, especially law enforcement. Discussions within the HCI community have centred in this second category, particularly on genetic data security and privacy concerns \cite{baig2020m,grandhi2022spit,king2019becoming}. King's study explored individual engagement with DTC-GT, applying social exchange theory to understand privacy expectations and motivations, and highlighted broader implications for societal genetic privacy \cite{king2019becoming}. Baig et al. reported findings from interviews with 27 Canadian users of DTC-GT\footnote{The authors predominantly used the term ``at-home DNA testing'' in their work. To ensure consistency, we will use ``DTC-GT'' throughout our paper.} services, highlighting inconsistent mental models, underestimation of privacy implications, and a desire for greater transparency and control over their data \cite{baig2020m}. Based on the above two studies, Grandhi et al. further discovered that individuals who had already used DTC-GT services generally reported lower privacy and security concerns compared to non-users \cite{grandhi2022spit}.  

Our study builds on these two categories to examine UK public attitudes toward consumer-driven PRS. Firstly, previous research predominantly focuses on the use of genetic information by medical professionals, such as healthcare providers and researchers. When these studies consider other institutions accessing genetic information, they typically regard them as secondary or unintended users. In contrast, our study equally considers both medical and non-medical users of genetic information. This reflects the practical reality that non-medical entities can now also collect and analyse genetic data, allowing us to compare public attitudes toward health-related and non-health-related applications. Secondly, our approach goes beyond extending existing literature by using provocative questions to elicit a wide range of perspectives on PRS. Participants critically reflected on hypothetical PRS scenarios, generating diverse viewpoints. Additionally, we utilised interactive storyboards with the ContraVision technique to create immersive scenarios simulating real-life PRS usage, provoking participants to have a deeper reflection and discussion. It allowed us to extend previous research by clearly distinguishing between three types of attitudes toward PRS: abstract beliefs (such as interest, value, and reliability) about the acceptability of actions enabled by PRS, concrete attitudes about the permissibility of specific PRS applications, and personal willingness to provide genetic information for PRS usage.

\section{Research Questions}
Our study was motivated by the following three research questions (RQ):

\begin{itemize}
    \item RQ1: What are the challenges and barriers perceived by participants of using PRS services?
    \item RQ2: What are the benefits perceived by participants of using PRS services?
    \item RQ3: How do people's experiences and stories influence their informed decisions about potentially using PRS services in the future?
\end{itemize}

\begin{figure*}

  \includegraphics[width=\textwidth]{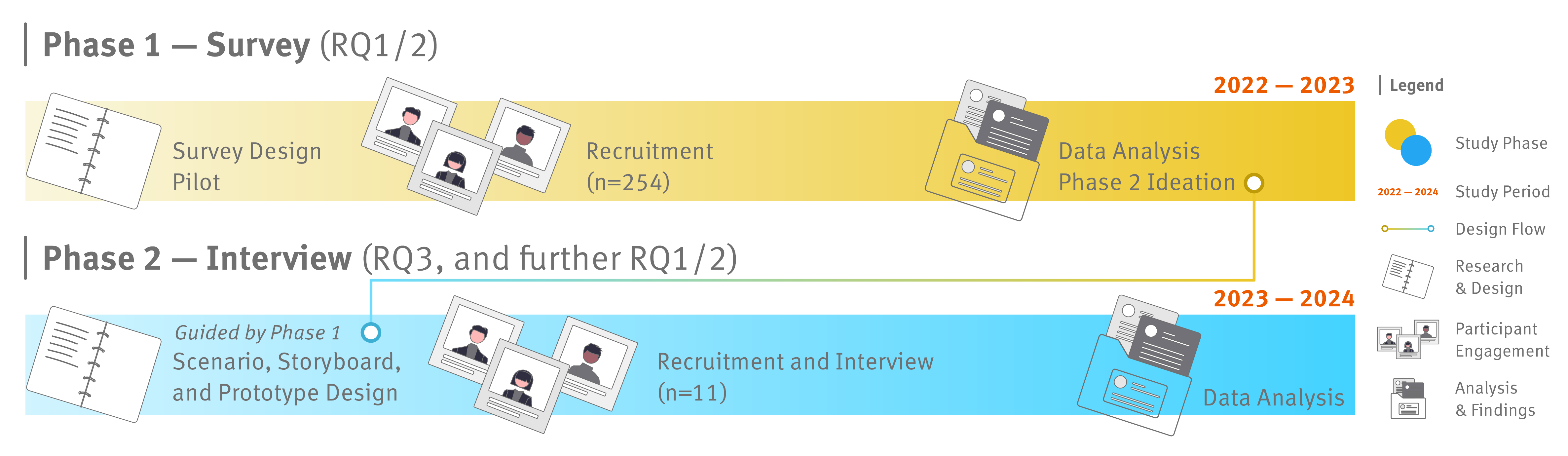}
  \caption{\textbf{Overview of Study Design: Two Sub-Studies Conducted.} A survey of 254 individuals explored public perception of PRS, followed by semi-structured interviews with 11 individuals for deeper insights. The findings from surveys guided the design and focus of interviews. A preliminary version of this work has also been reported in an extended abstract \cite{sun2024design}.}
  \Description{The figure illustrates two elongated panels representing the study design of two sub-studies. The first panel on the top represents Phase 1 of the study (2022-2023), which involves a survey aimed at answering Research Questions 1 and 2 (RQ1/2). This panel is divided into three sections: survey design, recruitment, and data analysis. The second panel on the bottom represents Phase 2 of the study (2023-2024), which includes interviews designed to address Research Question 3 (RQ3) and further explore RQ1/2. Similar to Phase 1, this panel is also divided into three sections: interview design, recruitment, and data analysis. An arrow connects the data analysis section of Phase 1 to the study design section of Phase 2, illustrating that the findings from Phase 1 guide the design and focus of Phase 2.}
  \label{fig:timeline}
\end{figure*}

To inquire into the three research questions, we conducted a mixed-methods study. As shown in Figure \ref{fig:timeline}, the study consists of two components, including the online survey (n=254) to address RQ1 and RQ2, then followed by the semi-structured interviews (n=11) to explore RQ3 and further investigate RQ1 and RQ2. For each sub-study, participants were provided with the foundational knowledge of PRS, including its benefits, limitations, and ethical considerations, to inform discussions. The survey results informed the design of materials for the subsequent interview. For clarity and coherence, we present the methods and findings of each sub-study together.

All study procedures were approved by the Edinburgh Medical School Research Ethics Committee (22-EMREC-040). All participants provided informed consent.

\section{Survey Study}

\subsection{Survey Methods and Protocol}

Our survey included both quantitative (scaled and binary questions) and qualitative (open-ended) questions. We designed the survey to encourage respondents to `diverge' rather than `converge' in their insights. To realise this, we included fewer objective questions and provided more opportunities for respondents to express their perspectives through open-ended prompts. We formulated a set of forward-looking but provocative questions to explore diverse perceptions of PRS applications. For example, we invited respondents to express their willingness (on a 5-Likert scale) to use personal DTC-GT and PRS results under five societally sensitive contexts, including school (under the UK context), insurance companies, embryo selection, sperm/egg donation, and dating apps. Schools and insurance companies could use PRS in decision-making that affects individuals' opportunities and access to services. In contrast, reproductive choices (embryo selection, sperm/egg donation) and dating scenarios directly involve personal decisions that could be influenced by genetic risk profiles. No questions in the survey were mandatory to answer, and respondents were allowed to skip questions or choose the ``prefer not to say'' option. The full contents of the survey can be accessed in Supplementary Material.

The survey (n=254) was open and disseminated online between 14 November 2022 and 31 March 2023. Respondents had to be at least 18 years old, fluent in English, and residing in the UK. Any professional or dependent relationship with researchers was grounds for exclusion. No prior knowledge of PRS was required to complete the survey, and we provided the lay summaries of PRS before posing the questions. Initially, the survey was disseminated through social media platforms, including LinkedIn, Twitter (now known as `X'), Facebook, and Reddit. Due to low participant engagement observed during the first month, commercial promotions were subsequently employed on Facebook and Reddit. No incentives were offered for survey participation. Ultimately, the survey was viewed 7298 times. Of the 260 responses, 254 were valid and analysed, while 6 were accessibility requests. The sample of 254 generated a 6.11\% margin of error with a 95\% confidence level. In other words, if we were to repeat the survey multiple times with different samples, 95\% of those surveys would have results within 6.11\% above or below the reported values.

Mixed methods were utilised for analysis. Quantitative responses, such as those from the 5-point Likert scale and categorical questions, were transformed into numerical values for statistical analysis, which led to the findings presented in Sections \ref{survey-value} and \ref{survey-interest}. As outlined by Braun and Clarke \cite{braun2012thematic,Braun2019}, we then thematically analysed (TA) 539 qualitative insights gathered from different open-ended survey questions to explore potential barriers to PRS usage, which informed Section \ref{survey-ten}. Given the varied nature of the questions, our approach to TA aimed to integrate diverse perspectives into a cohesive set of themes. We began with open coding, where each response was systematically coded to identify key ideas. Codes were generated at sentence-level and also, if appropriate, at the level of a whole response to a question. Through our coding of the data, we developed codes that represented both commonalities and unique viewpoints across different questions. We then employed axial coding to group these codes into broader categories, which enabled us to identify relationships between individual insights and consolidate them into ten overarching barriers to PRS usage. This analytic process provided a clear and structured understanding of participants' attitudes and concerns, which formed the foundation for subsequent interview design and exploration. Drawing on Bowman et al.'s discussion of TA in healthcare HCI at CHI \cite{bowman2023using}, we emphasised collaborative teamwork throughout the analysis process. Initial coding and generation of the axial codes and themes was led by the first author, with the second and third authors checking and supporting iterations of the analysis after each of these stages.
\label{survey-method}
\subsection{Survey Findings}

\aptLtoX{\begin{table}
\caption{\textbf{Demographics of Survey Respondents} (n=254).}
\label{tab:survey-demo}
\begin{tabular}{llrr}
\toprule
\textbf{Demographic} & \textbf{Response Options} & \textbf{\begin{tabular}[c]{@{}r@{}}Respondents\\ (n=254)\end{tabular}} & \textbf{Percentage} \\ \midrule
\multirow{7}{*}{Age} & 18-24 & 63 & 24.8\% \\
 & 25-34 & 83 & 32.7\%  \\
 & 35-44 & 49 & 19.3\%  \\
 & 45-54 & 23 & 9.06\%  \\
 & 55-64 & 28 & 11.0\%  \\
 & 65 and above & 7 & 2.76\%  \\
 & Prefer not to say / blank response & 1 & .394\%  \\ \midrule
\multirow{7}{*}{\begin{tabular}[c]{@{}l@{}}Gender\\ (multiple-choice)\end{tabular}} & Female & 118 & 46.5\%  \\
 & Male & 113 & 44.5\%  \\
 & Non-binary & 12 & 4.72\% \\
 & Transgender & 3 & 1.18\%  \\
 & Genderqueer & 3 & 1.18\%  \\
 & Other & 4 & 1.57\%  \\
 & Prefer not to say / blank response & 7 & 2.76\%  \\ \midrule
\multirow{6}{*}{Ethnicity} & White & 207 & 81.5\%  \\
 & Asian, or Asian British & 13 & 5.12\%  \\
 & Black, Black British, Caribbean, or African & 2 & .787\%  \\
 & Mixed or multiple ethnic groups & 15 & 5.91\%  \\
 & Other ethnic group & 6 & 2.36\%  \\
 & Prefer not to say / blank response & 11 & 4.33\%  \\ \midrule
\multirow{7}{*}{Education} & Primary & 0 & 0  \\
 & Secondary & 44 & 17.3\%  \\
 & Bachelors / Associate & 94 & 37.0\%  \\
 & Masters & 55 & 21.7\%  \\
 & PhD, or other advanced degrees (e.g., JD, MD, MBA) & 35 & 13.8\%  \\
 & Other & 16 & 6.30\%  \\
 & Prefer not to say / blank response & 10 & 3.94\%  \\ \midrule
\multirow{5}{*}{Location} & England & 154 & 60.6\%  \\
 & Scotland & 62 & 24.4\%  \\
 & Wales & 19 & 7.48\%  \\
 & Northern Ireland & 16 & 6.30\%  \\
 & Prefer not to say / blank response & 3 & 1.18\%  \\ \midrule
 \multirow{3}{*}{Experience} & DTC-GT & 43 & 16.9\%  \\
 & PRS & 15 & 5.91\%  \\
 & DTC-GT and PRS & 14 & 5.51\% \\
 \bottomrule
\end{tabular}
\end{table}}{\begin{table*}
\caption{\textbf{Demographics of Survey Respondents} (n=254).}
\label{tab:survey-demo}
\begin{tabular}{llrrl}
\toprule
\textbf{Demographic} & \textbf{Response Options} & \textbf{\begin{tabular}[c]{@{}r@{}}Respondents\\ (n=254)\end{tabular}} & \textbf{Percentage} &  \\ \midrule
\multirow{7}{*}{Age} & 18-24 & 63 & 24.8\% & \Chart{0.248} \\
 & 25-34 & 83 & 32.7\% & \Chart{0.327} \\
 & 35-44 & 49 & 19.3\% & \Chart{0.193} \\
 & 45-54 & 23 & 9.06\% & \Chart{0.0906} \\
 & 55-64 & 28 & 11.0\% & \Chart{0.110} \\
 & 65 and above & 7 & 2.76\% & \Chart{0.0276} \\
 & Prefer not to say / blank response & 1 & .394\% & \Chart{0.00394} \\ \midrule
\multirow{7}{*}{\begin{tabular}[c]{@{}l@{}}Gender\\ (multiple-choice)\end{tabular}} & Female & 118 & 46.5\% & \Chart{0.465} \\
 & Male & 113 & 44.5\% & \Chart{0.445} \\
 & Non-binary & 12 & 4.72\% & \Chart{0.0472} \\
 & Transgender & 3 & 1.18\% & \Chart{0.0118} \\
 & Genderqueer & 3 & 1.18\% & \Chart{0.0118} \\
 & Other & 4 & 1.57\% & \Chart{0.0157} \\
 & Prefer not to say / blank response & 7 & 2.76\% & \Chart{0.0276} \\ \midrule
\multirow{6}{*}{Ethnicity} & White & 207 & 81.5\% & \Chart{0.815} \\
 & Asian, or Asian British & 13 & 5.12\% & \Chart{0.0512} \\
 & Black, Black British, Caribbean, or African & 2 & .787\% & \Chart{0.00787} \\
 & Mixed or multiple ethnic groups & 15 & 5.91\% & \Chart{0.0591} \\
 & Other ethnic group & 6 & 2.36\% & \Chart{0.0236} \\
 & Prefer not to say / blank response & 11 & 4.33\% & \Chart{0.0433} \\ \midrule
\multirow{7}{*}{Education} & Primary & 0 & 0 & \Chart{0} \\
 & Secondary & 44 & 17.3\% & \Chart{0.173} \\
 & Bachelors / Associate & 94 & 37.0\% & \Chart{0.37} \\
 & Masters & 55 & 21.7\% & \Chart{0.217} \\
 & PhD, or other advanced degrees (e.g., JD, MD, MBA) & 35 & 13.8\% & \Chart{0.138} \\
 & Other & 16 & 6.30\% & \Chart{0.063} \\
 & Prefer not to say / blank response & 10 & 3.94\% & \Chart{0.0394} \\ \midrule
\multirow{5}{*}{Location} & England & 154 & 60.6\% & \Chart{0.606} \\
 & Scotland & 62 & 24.4\% & \Chart{0.244} \\
 & Wales & 19 & 7.48\% & \Chart{0.0748} \\
 & Northern Ireland & 16 & 6.30\% & \Chart{0.063} \\
 & Prefer not to say / blank response & 3 & 1.18\% & \Chart{0.0118} \\ \midrule
 \multirow{3}{*}{Experience} & DTC-GT & 43 & 16.9\% & \Chart{0.169} \\
 & PRS & 15 & 5.91\% & \Chart{0.059} \\
 & DTC-GT and PRS & 14 & 5.51\% & \Chart{0.0551} \\
 \bottomrule
\end{tabular}
\end{table*}}

Table \ref{tab:survey-demo} presents the demographics of the survey respondents, who ranged in age from 18 to 78 years, with an average age of 35.2 years. Among the respondents, 46.5\% were female, 81.5\% identified as White, and 60.6\% resided in England. 72.5\% of respondents had at least a Bachelor's or Associate-level education. Regarding experience, 16.9\% had used a DTC-GT service, and 5.91\% had generated a PRS result. Below, we present our survey findings in three sections.

\begin{figure*}
  \includegraphics[width=\textwidth]{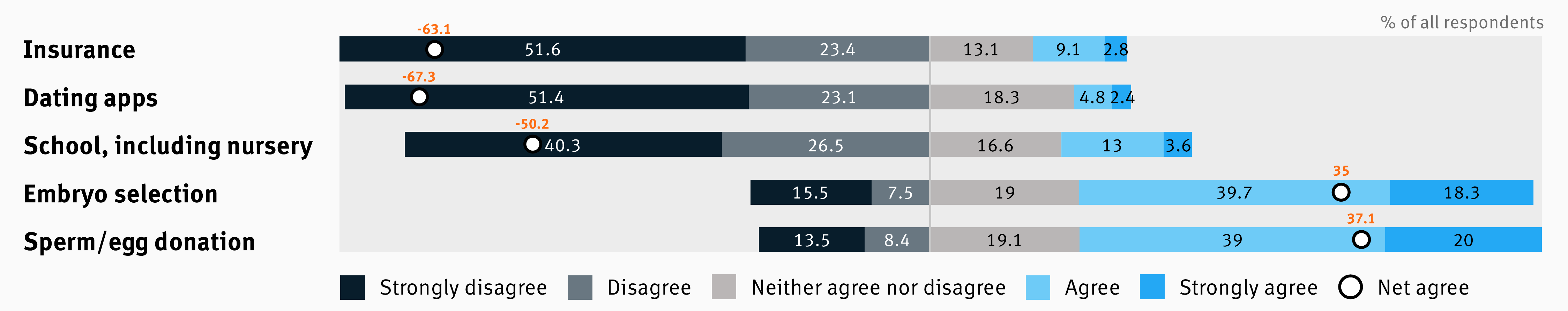}
  \caption{\textbf{Responses to the Question: ``To what extent do you agree the following institutions are being allowed to use personal DTC-GT \& PRS results in a commercial context, on the condition that you consent?''.} ``Net agree'' calculated as the total number of \textit{agreed} responses (``Agree'' and ``Strongly agree'') minus the total number of \textit{disagreed} responses (``Disgree'' and ``Strongly disagree'').}
  \Description{The figure shows the distribution of responses (on a 5-Likert scale) to a specific question across five different contexts: insurance, dating apps, school (including nursery), embryo selection, and sperm/egg donation. The responses are visualised to indicate the degree of agreement or disagreement with using DTC-GT and PRS results in these contexts. In the order presented - insurance, dating apps, school, embryo selection, and sperm/egg donation - there is a noticeable trend where respondents are more inclined to agree with the use of DTC-GT and PRS results in the latter contexts. The "net agree" values for the first three contexts (insurance, dating apps, and school) fall within the "strongly disagree" range. In contrast, the "net agree" values for the last two contexts (embryo selection and sperm/egg donation) are within the "agree" range.}
  \label{fig:survey-commercial}
\end{figure*}

\subsubsection{Perceived value, reliability and future applications towards PRS} Overall, respondents exhibited a neutral stance on their interest, perceived value, and reliability of PRS, with reliability receiving the lowest endorsement. We asked respondents to rate their willingness to use DTC-GT and PRS results in five sensitive contexts (schools, insurance, embryo selection, sperm/egg donation, and dating apps) on a 5-point Likert scale. As shown in Figure \ref{fig:survey-commercial}, respondents were more inclined to agree with DTC-GT and PRS use in embryo selection and sperm/egg donation, than in insurance, dating apps, or schools. Opinions on reproductive uses were slightly more polarised than in the other three contexts: while the majority of respondents ``agreed'' with the use of DTC-GT and PRS in embryo selection and sperm/egg donation, a similar number either ``strongly agreed'' or ``strongly disagreed'' with its use in these contexts. In contrast, the responses for using DTC-GT and PRS in insurance, dating apps, and schools were more uniformly distributed around the ``strongly disagreeing'' points.
\label{survey-value}

\begin{figure*}
  \includegraphics[width=\textwidth]{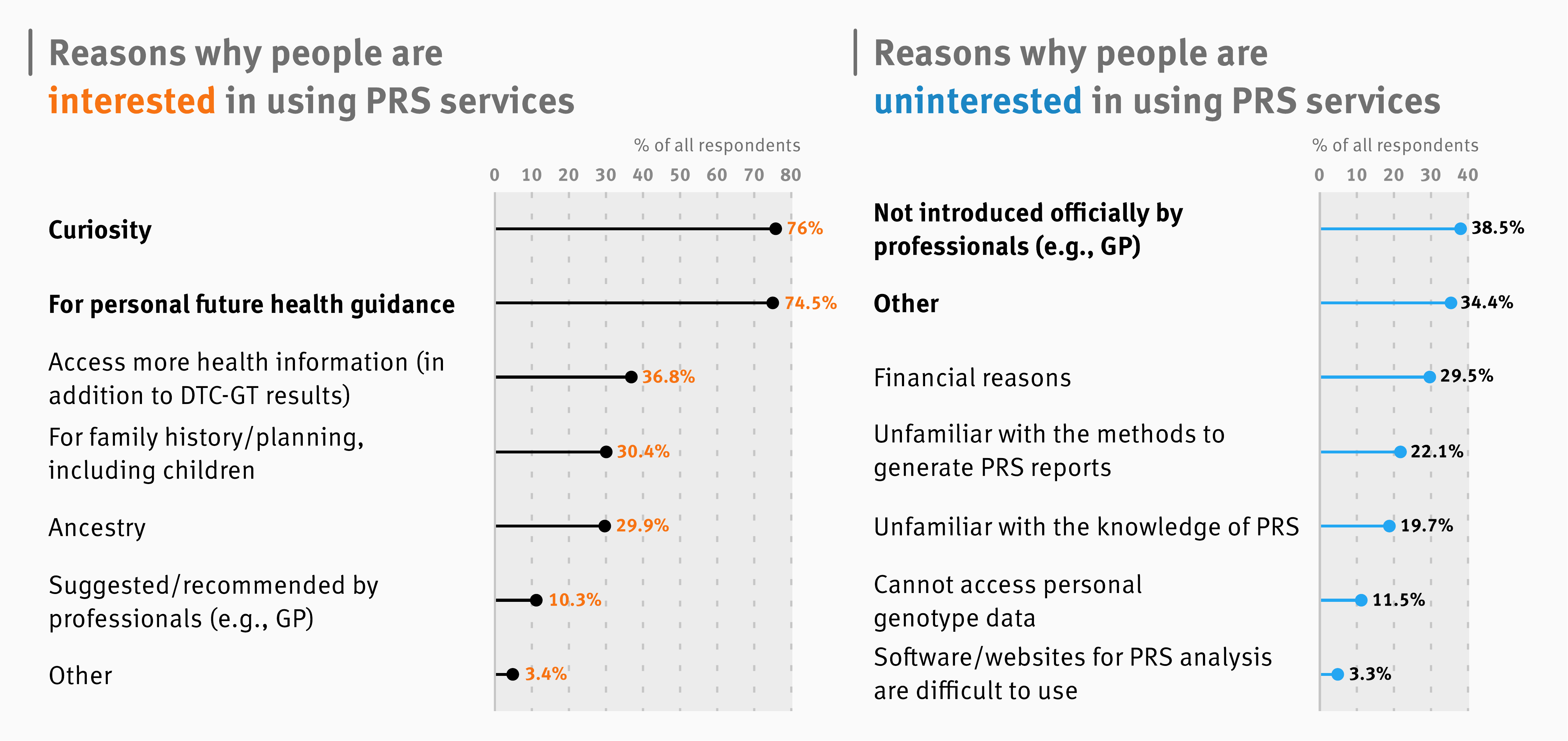}
  \caption{\textbf{Reasons of Interested and Uninterested in Using PRS Services.} Curiosity and desire for personal future health guidance as top motivators. Less structured reasons for not seeking PRS services, with the top reason being not introduced officially by professionals (e.g., GP).}
  \Description{The figure demonstrates the reasons for interest and disinterest in using PRS services. For interest in PRS services, reasons are presented in order: curiosity, personal future health guidance, access to more health information beyond DTC-GT results, family history and planning (including children), ancestry, suggestions or recommendations by professionals (e.g., GP), and other self-reported reasons. For disinterest in PRS services, reasons are shown in order: lack of official introduction by professionals (e.g., GP), other self-reported reasons, financial constraints, unfamiliarity with the methods for generating PRS reports, lack of knowledge about PRS, inability to access personal genotype data, and difficulty using software or websites for PRS analysis. Ranked from most to least chosen.}
  \label{fig:survey-motivations}
\end{figure*}

\subsubsection{Factors driving interest and uninterest in using PRS}
\label{survey-interest}
Figure \ref{fig:survey-motivations} (left) demonstrates that the motivations for seeking PRS vary (n=204). Curiosity (76\%) and personal health guidance (74.5\%) were the top reasons for seeking PRS. Other reasons for seeking PRS included the pursuit of more comprehensive health information (36.8\%), family history and planning considerations (30.4\%), ancestry information (29.9\%), and recommendations from healthcare professionals (10.3\%). As shown in Figure \ref{fig:survey-motivations} (right), the reasons for not seeking PRS (n=122) were less structured. They predominantly centred around the lack of professionals' (such as GP's) recommendations (38.5\%). Other reasons included financial constraints (29.5\%), unfamiliarity with the methods to generate PRS reports (22.1\%) and PRS knowledge (19.7\%), unable to access personal genotype data (11.5\%) and software/websites for PRS analysis are difficult to use (3.3\%). Self-filled other reasons (34.4\%), which were not on the list of options, were also noted. Key themes included scepticism about PRS, data privacy concerns, potential emotional reactions, perceived value, mistrust in providers, and doubts about genetics.

\begin{table*}
\caption{\textbf{Ten Barriers to PRS Usage}. The prompt questions accompanying the high-level themes, developed using key quotes, were designed to define their scope and provoke participants' perspectives.}
\label{tab:ten-barrier}
\begin{tabular}{p{0.8cm}p{4.2cm}p{9cm}} 
\toprule
\begin{tabular}[c]{@{}l@{}}\textbf{Barrier}\\ ($\oslash$)\end{tabular} & \begin{tabular}[c]{@{}l@{}}\textbf{High-level Theme} (abbreviation \\in bold) -- prompt questions\end{tabular} & \begin{tabular}[c]{@{}l@{}}\textbf{Key Quotes}, with respondents' index\end{tabular} \\\midrule
$\oslash$1               & Data \textbf{privacy} -- How secure is the PRS system in protecting your personal data?              & ``\textit{The only people who should know the details are yourself, and your doctor if you approve to share the information. Data privacy would not permit further.}'' (\texttt{R236})                                         \\ \midrule

$\oslash$2               & \textbf{Mental health} impact -- How might PRS results influence someone's emotional well-being or stress levels?               & ``\textit{I would not like to know if I am likely to get a condition such as Alzheimer's as I feel it would be like a ticking time bomb and very big worry if I were to find out if I were prone to a condition as such in the future and it would impact my quality of life and stress levels etc. in the meantime.}'' (\texttt{R138})                                        \\\midrule

$\oslash$3               & \textbf{Interpretation} by professionals -- Can doctors and health experts clearly explain PRS results to patients?                & ``\textit{Results are invaluable because they rely on interpretation from the company which may change over time or have a requirement for subtle understanding of complex interactions.}'' (\texttt{R250})                                       \\\midrule

$\oslash$4                & \textbf{Insurance} implications -- Could using PRS in insurance decisions be fair and beneficial?              &  ``\textit{Charging a person more because of genetic factors indicating risk, rather than aiding in better prevention pro-actively is a dangerous activity that will lead to genetics being viewed as more important than free will and prevention.}'' (\texttt{R204})                                        \\\midrule
$\oslash$5                & \textbf{Ageing}: benefit for the elderly -- How can PRS make a difference in senior citizens' health or care?              & ``\textit{I've reached an age where it is unlikely to provide useful health information.}'' (\texttt{R201})                                      \\\midrule
$\oslash$6                & Ethical considerations (\textbf{Ethics}) -- Are there other ethical concerns in how PRS is used or applied?               & ``\textit{I am not clear in my own mind about the ethics of the sperm/egg eugenics.}'' (\texttt{R61})                                        \\\midrule
$\oslash$7                & \textbf{School} -- Is it beneficial and safe to use PRS under the school contexts?                & ``\textit{... given the PRS is a chance of disease rather than a diagnosis, generally with a low likelihood, sharing with schools risks a `pre-crime' situation where children are treated as disabled or talented when they are not, reducing their outcomes and reducing their agency.}'' (\texttt{R61}) \\ \midrule
$\oslash$8                & \textbf{``Next steps''}: preventative methods -- Once you get your PRS results, what actions should you consider? & ``\textit{It could make me aware of some disease that is rare and has few treatments. I would prefer not to add that worry to my life when I have no control over it.}'' (\texttt{R208})                                       \\\midrule
$\oslash$9                & Education, \textbf{transparency and visualisation} -- How well does the PRS service explain its processes and findings to users?           & ``\textit{I feel I'd require a bit more background information initially and to understand how the company uses their science to extract the information.}'' (\texttt{R3})                             \\\midrule
$\oslash$10               & \textbf{Policy}, politics and \textbf{governance} -- What role do government guidelines or rules play in shaping PRS practices? & ``\textit{For treatments related to assisted conception this could improve chances of successful treatments, but I do think this should be financially regulated as it will likely be sold as another `add on' to IVF }[In vitro fertilisation] \textit{patients who are vulnerable to exploitation.}'' (\texttt{R32})                             \\ 
\bottomrule
\end{tabular}
\end{table*}

\subsubsection{Ten barriers to using PRS}
\label{survey-ten}

Table \ref{tab:ten-barrier} presents the ten-barrier themes we identified. They illuminate the intricate web of concerns, hopes, and understandings linked to PRS. To ensure an accurate interpretation of these high-level topics (e.g., privacy, mental health) and provoke deeper insights, we developed prompt questions for each theme. These questions provided context to refine the scope of these high-level themes and limit misinterpretations. The order of barriers does not imply ranking. Based on the ten barriers, we then designed ten corresponding stories under each topic with twenty endings for the interviews, to probe participants' insights and views. In the following contents, we use the symbol ``$\oslash$'' with the number (such as ``$\oslash$1'') to represent the corresponding barrier. We observed that, instead of barriers indicated by respondents being demonstrated independently, they were sometimes interlinked. For example, respondents were concerned about the mental health impact ($\oslash$2) of PRS, such as undue worries, due to the lack of effective preventative methods as the ``next step'' ($\oslash$8), after receiving their results. Another example is that respondents doubted healthcare professionals' interpretation of PRS results ($\oslash$3), driven by mistrust in the wider healthcare system and governance ($\oslash$10).

Overall, this survey of 254 respondents highlighted a nuanced landscape of participant attitudes towards PRS, with generally neutral yet context-dependent views. Acceptance was slightly polarised in reproductive scenarios, while insurance and educational settings elicited more consistently cautious responses. In exploring the motivations behind these attitudes, curiosity and the desire for personal future health guidance emerged as the primary drivers of interest in PRS, whereas reasons for disinterest were less structured, including factors like lack of professional recommendation and privacy concerns. Additionally, we identified ten barriers to using PRS, including ethical considerations, data privacy, and scepticism towards the reliability of professional interpretations. These findings underscored significant uncertainties and complex hesitations, necessitating further exploration to grasp their implications fully. Therefore, the subsequent interviews, building upon the survey findings, were designed to further unpack these barriers and motivations, providing a deeper understanding of participants' concerns, expectations, and perspectives regarding PRS.

\section{Interview Study}

\subsection{Interview Methods and Protocol}

\begin{figure*}
  \includegraphics[width=\textwidth]{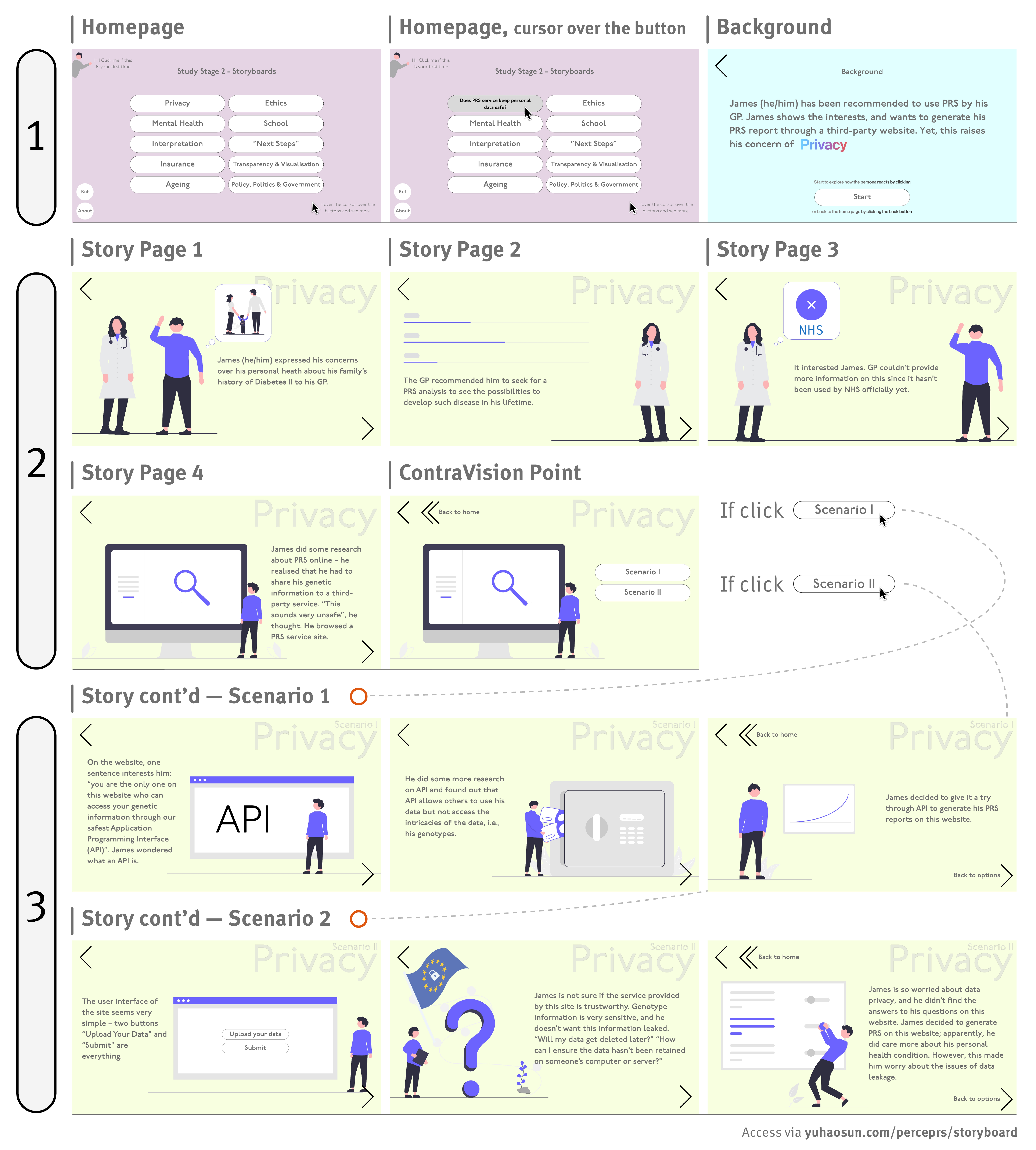}
  \caption{\textbf{The 138-Interface Interactive Storyboard Prototype: Illustrated Case on the Topic of ``Privacy'' ($\oslash$1).} The prototype is structured around story milestones and divided into three stages representing different contexts in the story. \textbf{Stage 1}: Participants visit the prototype site and select one of the ten story topics. Hovering over a button provides a further prompt question of the high-level topic. Upon selection, a background introduction to the story appears. \textbf{Stage 2}: The story begins by outlining the general topic and context. It then reaches the ContraVision point, where users must choose a scenario, leading to different story outcomes. \textbf{Stage 3}: The story continues based on the chosen scenario. At the end, participants have the option to return to the ContraVision point to select a different scenario and explore alternate story outcomes.}
  \Description{Same as the figure title.}
  \label{fig:isw}
\end{figure*}

Based on the ten barriers summarised in Table \ref{tab:ten-barrier} and the qualitative responses in the survey, we developed ten PRS user stories with vignettes, with two outcomes each. In the UK, attitudes towards genetics have become more positive nowadays, especially during the pandemic \cite{theGeneticsSociety-perception} -- which may have shaped participants' expectations and biases -- thus being influenced by preconceived narratives. To address this and go beyond the limitations of single-outcome stories, which can constrain exploration or reinforce preconceived biases, we employed the ContraVision technique, which presents contrasting (i.e., positive and negative) but open-ended outcome scenarios to foster critical reflection and balanced engagement \cite{mancini2010contravision}, into these ten stories. We then adapted ContraVision to deviate from its traditional use of explicitly binary framings and labels; instead, we presented outcomes without explicitly informing the sentiment polarity, to encourage participants to critically consider the range of implications -- positive, negative, or otherwise -- through their own subjective lens. To empower participants to control the pace of the stories, engage deeply with each scene to establish personal connections and reflections, and create a more dynamic, participatory, and `story-with' interview experience, we developed a 138-interface interactive storyboard prototype (access via \href{https://yuhaosun.com/perceprs/storyboard/}{https://yuhaosun.com/perceprs/storyboard/}) with Adobe XD to realise interactive digital storytelling, as Figure \ref{fig:isw} shown. Participants interacted with the prototype on a web browser. The summaries of ten stories can be found in Appendix \ref{sum-10story}. 

Before the interview, participants reviewed abstracts of all ten stories and selected three themes that they would like to read the full stories about and discuss further during the interview. At the start of the interview, participants were introduced to PRS, with emphasis on two alternative terms ``health prediction'' and ``genotype data'' to aid understanding. Followed by a brief chat, participants had three approximately 15-minute sessions, corresponding to three barrier themes they chose previously. In each session, participants were first invited to read the story with two outcome scenarios. After completing one scenario, participants had the option to return to the ContraVision point and select the alternative scenario. As participants engaged with the stories, some raised questions regarding the scenarios, either to clarify details or explore alternative interpretations. Upon completing the reading, participants were asked a series of questions based on the `funnel technique' \cite{rosala2022funnel}, starting with their overall impressions of the stories and progressively examining their personal, social, and community-level connections to them. During this process, participants frequently reflected and connected on how the personas and scenarios in the stories resonated -- or did not resonate -- with their own experiences or attitudes. At the end of the interview, participants were invited to rank the priority of the barrier themes they chose, providing brief reasons for their ranking. As more interviews were conducted over time, we evaluated the responses to our prompt questions and added new questions inspired by the past interviews. A full interview protocol and prompts can be accessed in Supplementary Material. 

We conducted eleven 1-to-1 semi-structured interviews (n=11) via Microsoft Teams between July 2023 and March 2024. Participants were recruited from various means: two from previous survey respondents, two from social media and public engagement networks, and seven from Prolific\footnote{\href{https://www.prolific.com/}{https://www.prolific.com/}}, an online participant pool \cite{palan2018prolific}. All participants had to be at least 18 years old and living in the UK. Participants were not required to have prior PRS knowledge; instead, they were invited to share their healthcare experiences before the interview, which informed the discussion. Participants received a \pounds10 gift card or platform token for their time and contribution. All interviews were audio-recorded and deleted after transcription. 

To analyse the interview data, we employed a similar TA process adopted in the survey sub-study, as detailed in Section \ref{survey-method}, combining open and axial coding to integrate diverse perspectives into cohesive themes. Building on the survey analysis, which primarily summarised and categorised responses from open-ended questions across a large dataset, the interview analysis aimed to probe deeper into participants' nuanced perceptions through a more interpretive, exploratory and dynamic TA approach. All authors actively contributed to coding, theme refinement, and critical discussions to ensure robustness and consistency in the final themes.

\subsection{Interview Findings}

\begin{table*}
\caption{\textbf{Demographics of Interview Participants} (n=11). (a) ``Health History'' includes mental health, physical health, or multimorbidity history. Multimorbidity indicates the presence of two or more long-term health conditions. (b) At the conclusion of each interview, participants ranked the priorities of the PRS barriers discussed; the order presented reflects the priorities determined by participants. Some participants discussed more than three barriers if time allowed after covering their top three choices. Table \ref{table:theme-vis} in Appendix \ref{app:theme-pri} visualised the barriers selected by all participants and their corresponding priorities.}
\label{tab:interview-demogra}
\begin{tabular}{p{0.5cm}p{0.5cm}p{1cm}p{1cm}p{1.5cm}p{1cm}p{2.8cm}p{4cm}} 
\toprule
\texttt{\textbf{PID}}  & \textbf{Age} & \textbf{Gender} & \textbf{Ethnicity} & \textbf{Education}           & \textbf{Location} & \textbf{Health History\textsuperscript{a}}                                                        & \textbf{Barriers Chosen}, ranked as their priorities\textsuperscript{b}                        \\ \midrule
\texttt{\textbf{P1}}    & 34  & Male   & Asian     & Masters             & Scotland & Psoriasis                                                             & ``Next Steps'' ($\oslash$8), Mental Health ($\oslash$2), Interpretation ($\oslash$3), Ageing ($\oslash$5)              \\
\texttt{\textbf{P2}}   & 25  & Female & White     & Masters             & Scotland & n/a                                                                   & Ethics ($\oslash$6), Interpretation ($\oslash$3), School ($\oslash$7)                                   \\
\texttt{\textbf{P3}}      & 39  & Female & White     & Bachelors / Associate & England  & Anxiety, Depression                                                   & ``Next Steps'' ($\oslash$8), Mental Health ($\oslash$2), Ethics ($\oslash$6)                              \\
\texttt{\textbf{P4}}    & 31  & Male   & Asian     & Bachelors / Associate & England  & Marfan Syndrome, Pelvic Pain, Neurogenic Bladder and Neurogenic Bowel & Ageing ($\oslash$5), Interpretation ($\oslash$3), ``Next Steps'' ($\oslash$8)                             \\
\texttt{\textbf{P5}}    & 49  & Female & White     & Bachelors / Associate & Scotland & n/a                                                                   & Ageing ($\oslash$5), Policy and Governance ($\oslash$10), Insurance ($\oslash$4), Mental Health ($\oslash$2)         \\
\texttt{\textbf{P6}}     & 36  & Female & White     & Masters             & England  & n/a                                                                   & ``Next Steps'' ($\oslash$8), Mental Health ($\oslash$2), Ageing ($\oslash$5)                             \\
\texttt{\textbf{P7}}   & 39  & Female & White     & Secondary           & Scotland & Anxiety, Depression                                                   & Privacy ($\oslash$1), Mental Health ($\oslash$2), Policy and Governance ($\oslash$10), Insurance ($\oslash$4), Ageing ($\oslash$5)\\
\texttt{\textbf{P8}}     & 53  & Female & White     & Masters             & England  & n/a                                                                   & Policy and Governance ($\oslash$10), Mental Health ($\oslash$2), Ethics ($\oslash$6)                     \\
\texttt{\textbf{P9}} & 31  & Male   & Black     & Masters             & England  & Sickle cell                                                           & Privacy ($\oslash$1), Mental Health ($\oslash$2), School ($\oslash$7)                                   \\
\texttt{\textbf{P10}}    & 30  & Male   & White     & Bachelors / Associate & England  & Anxiety, Depression, Temporomandibular Joint Dysfunction, Chronic pain, Trigeminal Neuralgia         & ``Next Steps'' ($\oslash$8), Mental Health ($\oslash$2), Ageing ($\oslash$5)                              \\
\texttt{\textbf{P11}}    & 51  & Male   & White     & Secondary           & England  & n/a                                                                   & Insurance ($\oslash$4), Transparency and Visualisation ($\oslash$9), Ageing ($\oslash$5)   \\
\bottomrule
\end{tabular}
\end{table*}

Table \ref{tab:interview-demogra} shows the demographics and the chosen themes of eleven interview participants. Participants were aged from 25 to 53, with an average age of 38 years. Six participants were female and five were male. Eight participants were White, two were Asian and one was Black. Nine participants had at least a Bachelor's or Associate-level education and two were educated in secondary level. Seven participants lived in England and four lived in Scotland. Six participants had at least one type of mental, physical health and/or multimorbidity history.

\begin{figure*}
  \includegraphics[width=\textwidth]{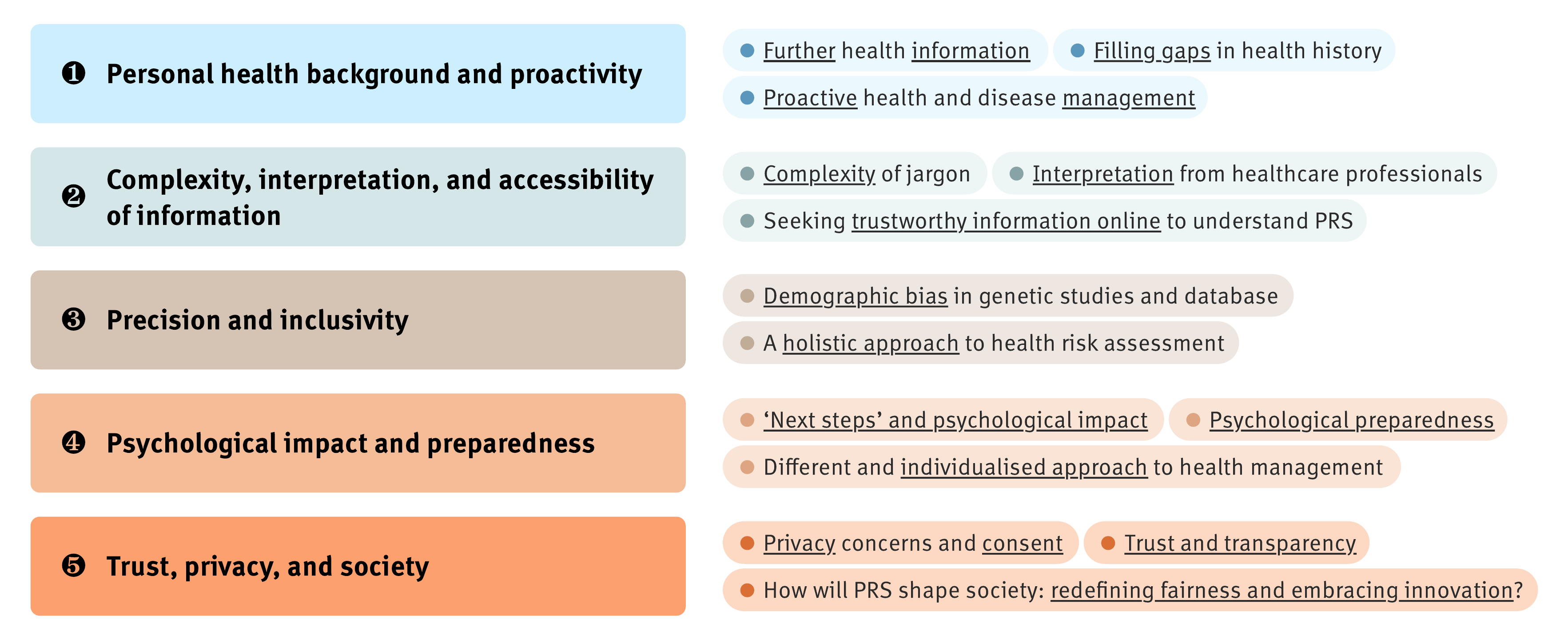}
  \caption{\textbf{Overview of Interview Findings.} We present our interview findings across five sections, with corresponding codes. Abbreviated codes are highlighted with an underline.}
  \Description{The figure shows five themes of interview findings with corresponding underlying codes, including (1) Personal health background and proactivity: Further health information, Proactive health and disease management, Filling gaps in health history; (2) Complexity, interpretation, and accessibility of information: Complexity of jargon, Interpretation from healthcare professionals, Seeking trustworthy information online to understand PRS; (3) Precision and inclusivity: Demographic bias in genetic studies and database, A holistic approach to health risk assessment; (4) Psychological impact and preparedness: `Next steps' and psychological impact, Psychological preparedness, Different and individualised approach to health management; (5) Trust, privacy, and society: Privacy concerns and consent, Trust and transparency, How will PRS shape society: redefining fairness and embracing innovation?}
  \label{fig:int-findings}
\end{figure*}

As Figure \ref{fig:int-findings} shown, we present our interview findings in five sections as the five following themes:

\begin{enumerate}
    \item \textbf{Personal health background and proactivity} -- participants viewed PRS as a proactive tool for understanding and improving their health, emphasising the importance of addressing personal health needs and filling gaps in health knowledge.
    \item \textbf{Complexity, interpretation, and accessibility of information} -- participants identified challenges in understanding complex PRS information and underscored the need to make it more accessible and meaningful.
    \item \textbf{Precision and inclusivity} -- participants highlighted the need for greater inclusivity in genetic studies and emphasised the value of a holistic approach to health risk assessment.
    \item \textbf{Psychological impact and preparedness} -- participants recognised the psychological impact of PRS and stressed the importance of preparedness and individualised approaches to health management.
    \item \textbf{Trust, privacy, and society} -- participants expressed concerns about privacy, trust, and societal implications related to PRS usage, ultimately reinforcing the importance of transparency and fair use of genetic information.
\end{enumerate}

\subsubsection{Personal health background and proactivity}

Participants viewed PRS as a means to enhance their personal agency in health management, whether by seeking further information, proactively managing health, or addressing gaps in their health history. This desire for control over health decisions was central to their interest in PRS.

\textbf{Further health information.} Some participants acknowledged PRS's potential to provide more health insights. \texttt{P1} emphasised the importance of prevention in health management, ``\textit{If I knew what I would get, then maybe there will be some kind of prevention... to avoid something that may trigger} [the disease].'' Similarly, \texttt{P11} hoped these services could provide him with some level of assistance that ``\textit{maybe}'' goes beyond ``\textit{reassurance},'' especially as he faced more health issues at the age of ``\textit{51} [year-old]'' and compared to ``\textit{somebody who is 21} [year-old]\textit{, who has probably less risk to health issues}.''

\textbf{Proactive health and disease management.} Additionally, participants reflect intrinsic motivations toward accessing PRS services. ``\textit{My own health is in my own hands...}'' \texttt{P8} said, is another theme throughout the participants. Immediately following this, \texttt{P8} expressed ``...\textit{I don't trust you} [healthcare system] \textit{and I will question you}.'' This scepticism and distrust of the healthcare system and proactive mindsets made them more inclined to use PRS to take control of their health decisions. Furthermore, \texttt{P11}, who believed ``\textit{plenty of people around... who aren't} [as healthy as \texttt{P11}],'' showed this proactivity through ``\textit{I cycle five or six times a week. I just enjoy being healthy and I enjoy being fit}.'' He also highlighted the importance of a positive mindset, ``[a] \textit{healthy lifestyle gives you a healthy mind}.'' Interestingly and similarly, \texttt{P6} passionately introduced her positive mindset and healthy lifestyle by doing ``\textit{meditation}'' and inspired by a book about ``\textit{hypnotherapy}'' -- those contents around ``\textit{spirits}.''

\textbf{Filling gaps in health history.} PRS services might be able to provide valuable health prediction information while enhancing users' knowledge and understanding, alleviating their uncertainty in medical interactions. \texttt{P11}, who was adopted as a child, has been feeling a persistent uncertainty about his medical history, which affects his interactions such as ``\textit{does cancer run in your family?}'' with healthcare professionals. His ``\textit{always}'' answer ``\textit{I don't know because I have no medical background, from the people, my birth parents so to speak, I had no knowledge of anything about them}'' made him ``\textit{find it a bit frustrating.}'' This sense of disconnect and insecurity in medical consultations is due to the lack of familial health information. PRS and genetic data, in this situation, fill a significant gap in his health history -- it represents an opportunity to gain insights into his genetic risks, something that he has been missing due to the absence of information about his biological family. \texttt{P11} admitted, this frustration ``\textit{is probably why I'm quite interested in things like this} [PRS] \textit{because I have no prior knowledge to ever go on.}''

\subsubsection{Complexity, interpretation, and accessibility of information}

Participants anticipated difficulties in accessing and understanding PRS due to complex jargon, reliance on healthcare professionals for interpretation, and unreliable online sources. These challenges reflect broader concerns about the accessibility of PRS information and its usability for non-experts.

\textbf{Complexity of jargon.} The `mystery' and complexity of PRS pose a challenge for potential users. Participants emphasised the importance of simplifying medical information for the general public who do not ``\textit{work with it} [PRS and its adjacent ones]'' in their everyday lives. Particularly, \texttt{P9} stressed the need for ``\textit{simplicity}'' in all relevant information, not just healthcare-related content. For example, in one of the stories, the Application Programming Interface (API) concept was suggested to address the PRS privacy issue ($\oslash$1). ``\textit{Some people don't know what API means,}'' said \texttt{P9}, ``[the service provider] \textit{needs to simplify it} [API]... \textit{rather than them} [people] \textit{going out to source for this information} [by] \textit{themselves}.'' Another theme of the educational and socio-economic barriers, that prevent individuals from fully engaging with and understanding the information related to PRS, was highlighted. \texttt{P8} said, ``\textit{also demographics as well... education... people who aren't quite so literate, they weren't understood. And they probably won't feel brave enough to question somebody who has letters after their name. So, if we need to make this} [PRS] \textit{accessible to everybody so that everybody feels able to say, I don't understand this} [PRS].''

\textbf{Interpretation from healthcare professionals.} Participants desired a healthcare approach that provides clear, precise information, and comprehensive support -- though PRS will not necessarily be provided by the healthcare system. This includes addressing the emotional and ethical aspects of PRS, highlighting the importance of both objective data and holistic care. \texttt{P4}'s frustration, generated by the ``\textit{time-consuming}'' and the uncertainty about the ``\textit{right path}'' to explore the answers, reflects a broader systemic issue where ``\textit{every patient goes through}'' the process of being left alone to interpret complex medical information. It eventually leaves patients in a state of confusion and increases the time and effort required to understand their health conditions accurately. When discussing a scenario where patients receive PRS results without proper guidance for interpretation, he showed his concerns towards ``\textit{transparency and accountability on parts of GPs.}'' Similarly, \texttt{P8} expressed this sentiment by noting that ``\textit{a lot of people don't have time to research}'' and ``\textit{they just take the doctor's words as law.}'' Furthermore, \texttt{P11}'s preference for technology-based information reveals a need for precise and objective medical data, free from human emotional influence. He described technology as ``\textit{very black and white}'' and has ``\textit{no emotions.}'' ``\textit{If you take the human element out of there, your results quite often are more precise,}'' he said. More specifically, ``\textit{if you strip that bit} [emotions] \textit{out, you're just left with fact.}'' Additionally, the necessity of a multidisciplinary approach to PRS management, addressing the clinical, emotional, and ethical dimensions, was mentioned. \texttt{P5} expected an interdisciplinary team there -- ``\textit{an assembler}'', and ``\textit{not just all clinical.}'' ``\textit{Both of them} [two scenarios in $\oslash$10] \textit{have a team involved to cover different aspects, like the therapy, and the ethical and clinical} [issues].'' She noted that ``\textit{the emotional face}'' is as important as ``\textit{the medical side of it }[PRS].''

\textbf{Seeking trustworthy information online to understand PRS.} When faced with the challenge of understanding their PRS, many participants turned to online resources as a primary means of gaining insights. In the mental health scenario ($\oslash$2), the main character, struggling to understand their PRS, reached out to a friend working in genetics for clarity. In response, \texttt{P3} mentioned that ``\textit{my mom would have nobody in her circle to talk to, she doesn't know anybody with that kind of knowledge} [PRS].'' Thus, \texttt{P3}'s mother needed to be ``\textit{relying on professionals or the internet.}'' However, professional information on the internet, such as ``[academic] \textit{papers},'' is usually ``\textit{hard to comprehend}'' and/or ``\textit{behind paywalls.}'' \texttt{P6}, who often and ``\textit{only}'' used a free database named ``\textit{PubMed}'' for healthcare information, shared her potential journey if she would like to find ``\textit{good results}'' about more PRS-related information. However, not everyone had access to or the ability to navigate such specialised databases like \texttt{P6}. In the end, although emerging LLM tools such as ``\textit{ChatGPT}'' were mentioned, participants expressed scepticism about the reliability and accuracy of such AI-driven tools when it comes to interpreting complex PRS data.

\subsubsection{Precision and inclusivity}

Concerns about the accuracy and inclusivity of PRS were driven by demographic biases in genetic studies. Participants called for a more holistic health assessment, combining genetic and non-genetic factors, to ensure more precise and equitable risk evaluations.

\textbf{Demographic bias in genetic studies and database.} The current genetic databases predominantly reflect White populations, leading to limitations in PRS results for diverse ethnic groups. \texttt{P8} was aware of this inherent demographic bias in genetic research and questioned: ``\textit{how wide the} [PRS] \textit{results are.}'' She said, ``\textit{from my understanding, they're mainly based on a Caucasian}\footnote{The term ``Caucasian'' was used by the participant, but we acknowledge this language is considered obsolete and not inclusive.} \textit{population, and they're very limited... it depends on which area of the population you're dealing with.}'' This bias, consequently, leads to lower accuracy and applicability of PRS for individuals from diverse ethnic backgrounds besides White. More inclusive genetic databases are needed to provide reliable PRS for all demographic groups. \texttt{P3} expressed a fundamental bias going beyond the genetic data itself ``\textit{...in medical teaching, they'll talk about how conditions will present usually in white men... I have seen medical instruction materials that will talk about like skin conditions and how they manifest... you need to be able to spot like a meningitis rash on a black person... it's not good enough to just keep using the same, the same shade, same body type, same sex over and over again.}''

\textbf{A holistic approach to health risk assessment.} The necessity of a holistic approach in risk assessment had been addressed by participants -- considering environmental, lifestyle, and other personal factors that influence health. \texttt{P5} hoped that the PRS would be ``\textit{done all the fact-finding about the individuals,}'' especially ``\textit{the other factors that play a role} [in risk assessments].'' A comprehensive understanding of an individual's background is crucial for accurate and meaningful -- risk assessment -- results.

\subsubsection{Psychological impact and preparedness}

Participants were concerned about the psychological impact of PRS, particularly the uncertainty surrounding the next steps after receiving results. They emphasised that different types of support or approaches to these next steps could result in varying levels of psychological preparedness.

\textbf{`Next steps' and psychological impact.} The potential burden of knowing about the high likelihood of developing a certain disease was highlighted. Several participants showed this sentiment, generated from the uncertainty of the `next steps.' \texttt{P11} recognised the emotional complexity and potential anxiety that comes with PRS. He suggested that the awareness of an imminent health issue may not always lead to beneficial outcomes, ``\textit{if somebody says:} [\texttt{P11}'s name]\textit{, there's going to be an 80\% chance that you're gonna develop Parkinson's in the next, I don't know, five years or something. I don't know how that would affect my life. I don't know if it would have a negative or positive effect on it. I'm just trying to work that out in my own mind... sometimes, I think it's better not to know. Because I can't change it... I could probably do something, but my understanding is I can't stop... like Parkinson's come in.}'' In contrast, this sentiment dramatically changed when discussing different diseases that PRS focuses on. In terms of Type 2 diabetes, \texttt{P11} said ``\textit{I'd be confident that I could change it.}'' A deeper concern about the psychological impact of knowing PRS was mentioned. \texttt{P5} suggested that for individuals prone to be ``\textit{worriers}'' or ``\textit{over-thinkers,}'' the mere possibility of a future health issue can lead to a ``\textit{snowball effect}'' of anxiety and stress. This continuous cycle of worry can significantly deteriorate their mental health. Consequently, this consideration forced participants to reflect on whether the benefits of knowing their PRS outweigh the potential psychological harms. Consistent with the nature of PRS, ``\textit{it's }[PRS] \textit{still just a possibility, it's not a guarantee,}'' \texttt{P5} noted.

\textbf{Psychological preparedness.} Despite the challenges, some participants highlighted the importance of psychological preparedness and the value of PRS in providing a sense of security and readiness to face future challenges. \texttt{P4}, who suffered from Marfan Syndrome, expressed that ``\textit{PRS is quite helpful}'' and enabled ``\textit{a sense of security}'' for himself. ``\textit{After getting a genetic opinion} [about the disease] \textit{I was diagnosed... it helped me prepare mentally} [for what] \textit{will happen in the future,}'' he noted. Mental strength also plays an important role. \texttt{P9}, who suffered from sickle cell disease, described himself as ``\textit{mentally strong.}'' He said, ``\textit{not everybody that has the sickness that I have survives. Even those} [who] \textit{survive,} [they] \textit{would be depressed.}'' \texttt{P11}, however, offered a contrasting view, highlighting that sometimes not knowing might be better for maintaining a positive outlook. He said ``\textit{I really believe in the power of positive thinking. Not to the point where you can positively think things away, but I don't think having the knowledge of an impending illness hanging over you would really be a positive thing for me personally in my life.}'' The power of community such as ``\textit{Macmillan}\footnote{\href{https://www.macmillan.org.uk/}{https://www.macmillan.org.uk/}},'' especially with the ``\textit{people that have a similar disease,}'' was also emphasised to go through this psychological preparedness. \texttt{P3} mentioned that ``\textit{...nobody else in that room is gonna fix your situation, but just airing out the things that you're worrying about and thinking about can be so helpful.}''

\textbf{Different and individualised approach to health management.} In two scenarios of the `next steps' story ($\oslash$8), we designed two different approaches as preventative methods after receiving PRS results -- changing lifestyle or taking corresponding medication -- to achieve the goal of prevention. Participants expressed the need for personalised health strategies that align with individual preferences, psychological dispositions, and willingness to adapt over time. \texttt{P10} showed his expectations of certainty, control and straightforwardness in the prevention process, through his preference of taking the medication which indicated to have a 99\% success rate of prevention. ``\textit{It's gonna do that} [prevention] \textit{with certainty,}'' he described it as ``\textit{a very sort of fulfilling step}'' and ``\textit{a very simple step forward.}''  The simplicity in action, taking a pill, is a minimal addition to daily routine compared to broader lifestyle changes, ``\textit{with the certainty that you're gonna be fine for that issue in the future,} [taking a pill] \textit{isn't really a big cost to take.}'' Then, \texttt{P10} described the lifestyle changes as ``\textit{a little bit more open-ended}'' with ``\textit{no certainties.}'' He said, ``\textit{you feel like you're doing something on goodwill, but you don't know.}'' \texttt{P3} discussed, ``\textit{what adults don't know?}'' In contrast, \texttt{P6} expressed a different narrative, favouring lifestyle changes over medication as her initial step -- leaning toward natural and holistic health management strategies. She preferred to ``\textit{monitor}'' the effectiveness of lifestyle changes and consider medication only if these measures prove insufficient over time, such as ``\textit{a year or two years.}'' \texttt{P10} and \texttt{P6} jointly demonstrated that the appropriate next step can vary greatly among individuals, suggesting a tailored approach to health interventions. The perceived psychological comfort provided by certain interventions plays a crucial role in decision-making. For \texttt{P10}, the certainty of a medical solution alleviates anxiety, while \texttt{P6} finds reassurance in the ability to monitor and adapt her health management strategy over time.

\subsubsection{Privacy, trust, and society}

Privacy and informed consent were critical concerns, closely tied to broader issues of trust and transparency in PRS. Participants also reflected on how PRS could reshape societal fairness, questioning its impact on equity and the social consequences of genetic innovation.

\textbf{Privacy concerns and consent.} Like other data-driven technologies, PRS services face tensions between stringent privacy protections and leveraging healthcare advancements. \texttt{P9} stressed, ``\textit{privacy should be the number one priority.}'' He further explained that ``\textit{without people giving you their consents, you can't do anything, you can't get any data.}'' \texttt{P9} then highlighted the issue of lengthy terms and conditions in the consent process. By pointing out that ``\textit{most people don't read}'' these documents and simply ``\textit{click `I Agree' and continue,}'' he indicated a need for more transparent and concise communication about privacy policies to ensure informed consent.

\textbf{Trust and transparency.} As we mentioned earlier, PRS will not necessarily be provided by the healthcare system. However, given that PRS is a component of the healthcare domain, participants sometimes associated it with the broader healthcare system. Trust in the healthcare system, thus, was a recurring theme, with breaches of confidentiality and perceived failures in care leading to significant distrust. \texttt{P7}'s lack of trust in the healthcare system was rooted in a personal experience where her friend's employment at her GP practice led to a breach of family privacy. \texttt{P7} shared, ``\textit{she} [\texttt{P7}'s friend] \textit{was asking me and telling me information about other family members. Although we are friends, I just was really concerned when there were other people in the room.}'' The vulnerability patients feel regarding their personal information and the potential for misuse, even among acquaintances. Similarly, \texttt{P11}'s mistrust stems from a ``\textit{terrible}'' experience with his father's GP practice, which ``\textit{let him} [\texttt{P11}'s father] \textit{down.}'' \texttt{P11} felt failed in their duty of care and eventually caused the loss of his father. The subsequent ``\textit{letter of an apology,}'' which contained inaccuracies and things ``\textit{were not true,}'' further eroded his trust. ``\textit{Nothing will change that situation. I can't bring my dad back,}'' \texttt{P11} said, ``\textit{I don't really have the time nor the inclination to fight them because I won't win.}'' Mishandling patient care and communication can lead to a breakdown in trust, potentially spreading to other healthcare technologies and services.

\textbf{How will PRS shape society: redefining fairness and embracing innovation?} We designed two possibilities for using PRS in the insurance industry ($\oslash4$). Scenario one offers the same price for different services to customers with varying PRS. Scenario two adjusts insurance fees based on risk levels: lower PRS individuals pay less and higher PRS individuals pay more. \texttt{P11}, describing himself as ``\textit{not cynical,}'' believed that those who are ``\textit{using the service more} [they] \textit{should pay more.}'' He expressed scepticism about a uniform pricing model in insurance, where everyone pays the same price. He argued that in such a system, infrequent users of healthcare services end up ``\textit{compensating}'' those who ``\textit{are draining the system.}'' This perceived inequity suggests a concern that uniform pricing fails to account for individual differences in healthcare utilisation, leading to an unfair financial burden on those who use fewer services. On the contrary, \texttt{P2} criticised the idea of ``\textit{penalising}'' individuals based on ``\textit{something that there is actually not really affecting you right now,}'' calling it unfair. Furthermore, \texttt{P2} pointed out that ``\textit{in the UK, we don't really need to think about it} [insurance] \textit{so much, because our healthcare is not dependent on insurance.}'' It contrasts with insurance-based systems in other countries, influencing her views on fairness and payment models. Although \texttt{P11} was concerned with the fairness of cost distribution based on service usage, while \texttt{P2} focused on the fairness of penalising individuals for potential risks rather than current health status -- they both revolved around fairness and equity in healthcare payment issues.

Overall, building on the survey findings, the interviews with 11 participants offered personal insights into the perceived challenges and benefits of PRS. Our participants engaged with the ContraVision-based PRS storyboards by relating them to their past experiences and stories, enabling them to reflect on PRS as an emerging and unfamiliar technology. The interview findings highlighted five themes including proactive health management, challenges in understanding genetic information, the need for inclusivity, psychological impacts, and societal concerns -- emphasising making genetic insights accessible and trustworthy, adopting a holistic approach to health, and ensuring transparency in data use.

\section{Discussion}

Our survey and interview studies jointly exhibit a nuanced landscape of public perceptions towards PRS, characterised by mixed feelings and, in some cases, ambivalence. Participants generally expressed a neutral stance, reflecting uncertainties driven by limited understanding and the early stage of PRS adoption. While some recognised the potential benefits of proactive health management, significant barriers were also highlighted, including psychological impacts, the complexity of genetic information, and concerns about privacy and data bias. Despite these challenges, participants acknowledged the potential of PRS to support personalised health insights, facilitate healthcare conversations, and promote preventive measures. In the following discussion, we revisit the interconnected benefits and barriers of PRS, examining how these dualities might shape future development and implementation strategies. We also consider the socio-technical challenges that may prevent PRS from being fully realised as a comprehensive system. We conclude the discussion by suggesting design implications to foster responsible PRS services that address public concerns, build trust, and ensure inclusivity in diverse contexts.

\subsection{Intertwined Perceived Benefits and Barriers to PRS}

In prior studies, the perceived benefits and barriers of PRS and broader genetic data have often been discussed separately as distinct aspects \cite{peck2022people,zhang2021public,baig2020m}. Additionally, some studies have provided valuable insights into critical issues brought by PRS, including data privacy \cite{king2019becoming,grandhi2022spit}, exacerbated health disparities \cite{kachuri2024principles,martin2019clinical}, possible psychological impacts \cite{peck2022people}, and potential misuses \cite{turley2021problems,yanes2024future,schoeler2019multi,selzam2017genome}. Building upon this foundation, \textbf{our study observed how the perceived benefits and barriers of PRS might co-exist and influence one another, demonstrating an intricate and ambiguous relationship between these aspects.} These tensions emerged not only within individual participants but also as contrasting perspectives across the participant pool. For instance, participants often expressed that the same aspect of PRS could be both beneficial and problematic, depending on the context. It illustrates the dual nature of PRS perceptions -- benefits and barriers are not mutually exclusive but part of a dialectical relationship. Participants simultaneously recognised the advantages of PRS, such as personalised health insights, while being concerned about potential drawbacks such as the psychological impacts of such insights and the uncertainty of what to do with them. We acknowledge the \textbf{complexity} of PRS, but its current structure, lacking foundational principles, remains \textbf{complicated} for individuals to fully grasp -- ultimately leading to \textbf{confusion}. This ambivalence indicates that efforts to promote PRS should address both aspects in an integrated manner, acknowledging the complexity of users' evaluations. Overall, we observed them coexisting in a nuanced and intertwined relationship, reflecting participants' complex assessments and hesitations. To further understand these ambivalent perceptions, in the next section, we broaden our perspective to discuss PRS as a complex socio-technical system, considering factors beyond the individual user level.

\subsection{Understanding PRS as a Complex Socio-Technical System}

As discussed previously, our findings observed an interplay between perceived benefits and barriers of PRS, highlighting both technical and non-technical dimensions. Drawing on socio-technical systems theory \cite{trist1981evolution,baxter2011socio}, these intertwined perceptions indicate that it may be more appropriate to view PRS not solely as a technical tool, which is what dominates much research and development surrounding PRS thus far, but as a component that should be understood, designed, and implemented within the broader context of \textbf{a complex socio-technical system}. To address this, we discuss the specific dimensions of complexity associated with PRS below.

\subsubsection{Non-Linearity and Probabilistic Nature}
Building upon the interconnected concerns, one key aspect of PRS complexity is its non-linear interactions and probabilistic nature. Genetic information interacts with environmental factors and lifestyle behaviours in unpredictable ways \cite{kachuri2024principles}, such as a genetic variant increasing risk only when paired with specific diets or toxin exposures, also known as ``gene-environment interaction'' \cite{virolainen2023gene}. It can cause small changes to have disproportionately large effects, complicating outcome predictions. As previously introduced, PRS provides probabilistic risk estimates, reflecting likelihoods rather than certainties, which challenges individuals in interpreting results \cite{torkamani2018personal}. Our findings, consistent with prior studies \cite{carleton2016fear,han2011varieties}, show participants' confusion and anxiety when probabilistic information conflicted with their expectations for clear cause-and-effect explanations. Our participants, moreover, voiced a desire for tools that contextualise probabilities within their personal circumstances. These findings extend the need for decision-support tools or systems that help individuals make sense of probabilistic information and navigate its implications for their lives. While our findings highlight the potential of AI-driven tools -- such as LLMs -- for interpreting PRS results, further evaluation is required to ensure they support user understanding, transparency, and ethical standards.

\subsubsection{Delayed and Indirect Feedback Loops}

Another significant challenge is the delayed and indirect feedback loops inherent in PRS. Unlike interventions that yield immediate outcomes, the effects of using genetic data \cite{manolio2013implementing} or making lifestyle adjustments \cite{diabetes2002reduction} may take years to manifest. For example, an individual informed by their PRS of an increased risk for type 2 diabetes may adopt dietary changes and increase physical activity, yet not see measurable health improvements for an extended period. Our findings highlight that participants frequently expressed frustration with the delayed feedback inherent in PRS -- lacking the ``next steps'' -- led to doubts about its effectiveness and diminished motivation to adhere to recommended actions. Prior research aligns with our findings, suggesting that timely feedback is crucial for maintaining user engagement and promoting behaviour change \cite{norman2013design,consolvo2008activity}, while delayed feedback can lead to decreased motivation and even abandonment of the associated systems \cite{goncalves2015motivating,pinder2018digital}. Based on these findings, we suggest designing PRS systems that integrate interim feedback mechanisms, such as progress tracking or predictive visualisations, to help users perceive incremental benefits and maintain long-term motivation.

\subsubsection{Risks of Data Governance and Commercialisation}

Our findings highlight data privacy as a central concern for participants, particularly regarding the long-term handling and security of personal genetic information by potential private PRS providers. Challenges in data governance are evident in DTC-GT services, as seen in 23andMe's data breaches despite measures to enhance security and trust \cite{23andme-actionplan:online,23andMeU64:online}. If 23andMe ceased operations, the handling of vast user data highlights the risks of inadequate oversight \cite{23andMe-brink:online}. A similar trend can be observed with the PRS analysis platform \textit{impute.me}, originally an open-source, non-profit initiative \cite{folkersen2020impute}. Among our 15 survey respondents who had generated a PRS report\footnote{See the ``Experience'' row in Table \ref{tab:survey-demo}.}, six used \textit{impute.me}. In 2022, the website was taken down \cite{impute-wiki:online} and now redirects to the website of Nucleus\footnote{\href{https://mynucleus.com/}{https://mynucleus.com/}}, a U.S. DTC-GT company founded in 2023. This transition to commercialisation uncovers the tension between user privacy and profit motives, where conflicts inevitably arise between safeguarding data and pursuing commercial gains. 

In addition to privacy issues, meanwhile, our findings point out that commercialisation may lead to PRS being offered as out-of-pocket services, thus limiting access to higher-income individuals and potentially exacerbating existing health inequities. Martin et al. have already highlighted that the clinical use of current PRS may exacerbate health disparities from a diversity perspective \cite{martin2019clinical}. We extend this concern to the commercial realm, as the pricing models adopted by DTC-GT companies like Nucleus can create significant barriers to access. As of 30 January 2025, Nucleus charges a \$39 annual membership fee for users who have a DNA file\footnote{Users who do not have their DNA files need first to obtain Nucleus' whole-genome sequencing for \$399 or choose from other DTC-GT services.}, allowing them to upload the file to Nucleus and get ``all of Nucleus' health reports'' which ``will update automatically to reflect your current health with new research, analyses, and changes to your lifestyle factors''\footnote{\href{https://mynucleus.com/faq}{https://mynucleus.com/faq}}. In contrast, the predecessor of Nucleus, \textit{impute.me}, previously provided PRS analysis for free, democratising access and enabling a broader demographic to use genetic insights without financial barriers. The transition from free, open-source initiatives to commercial entities with substantial fees exacerbates the digital divide \cite{van2017digital} and shows how commercialisation restricts equitable access to PRS. More worryingly, when paid services replace free platforms with different privacy policies, economically disadvantaged users are left without affordable options, exacerbating health inequities and undermining public trust in genetic testing services. Our findings suggest that neglecting PRS' socio-technical implications risks commercial interests overriding ethical responsibilities, leading to misuse or inequitable access. Therefore, we call for robust governance frameworks that prioritise transparency, long-term data security, and equitable access. Importantly, integrating PRS into covered healthcare expenses could help ensure that PRS is accessible to diverse populations, promoting health equity, and ultimately benefiting all segments of society.

\subsubsection{Challenges in Regulating Genetic Data Misuse}

We concluded our interview findings with a thought-provoking question: how will PRS shape society? Participants' reflections highlighted broader societal implications, prompting us to extend our discussion to a critical challenge: the regulation of genetic data misuse. Concerns about genetic discrimination are longstanding. In the U.S., before the Genetic Information Nondiscrimination Act (GINA) in 2008, individuals in the U.S. faced employment and insurance denial based on genetic information \cite{hudson2008keeping}. However, GINA has limitations, excluding life, disability, and long-term care insurance, leaving gaps for potential discrimination \cite{rothstein2009gina}. In countries without such legislative protections, the fears about the misuse of data by insurers \cite{muller2024uninsurable}, school \cite{kid-school:online} or employers \cite{chapman2020genetic}, along with the ethical implications of profiling based on genetic predispositions, are more pronounced and amplified by the delayed nature of potential consequences. Most importantly, PRS relies on probabilistic risk estimates instead of definitive results, complicating how laws define, classify, and protect individuals against genetic discrimination. Thus, potential questions arise about when and how such predictions become legally relevant, what constitutes discrimination or unfair treatment based on probabilistic data, and how to ensure that evolving algorithms and reference databases do not introduce new forms of inequity. We advocate for legal frameworks and transparent policies to safeguard individuals' rights. Equally crucial is fostering public trust through education and open dialogue, particularly in diverse cultural contexts where perceptions of PRS and genetic data may vary widely.

In summary, we examined PRS as a complex socio-technical system and unpacked key challenges, including its probabilistic nature, delayed feedback, data governance and commercialisation risks, and regulatory gaps. These challenges call for the need for holistic and collaborative solutions. In the next section, we propose design implications to address these complexities and aim to guide the responsible PRS ultimately.

\subsection{Designing for Impact: Implications, Responsibility, and Stakeholders in PRS}

\begin{figure*}
  \includegraphics[width=0.65\textwidth]{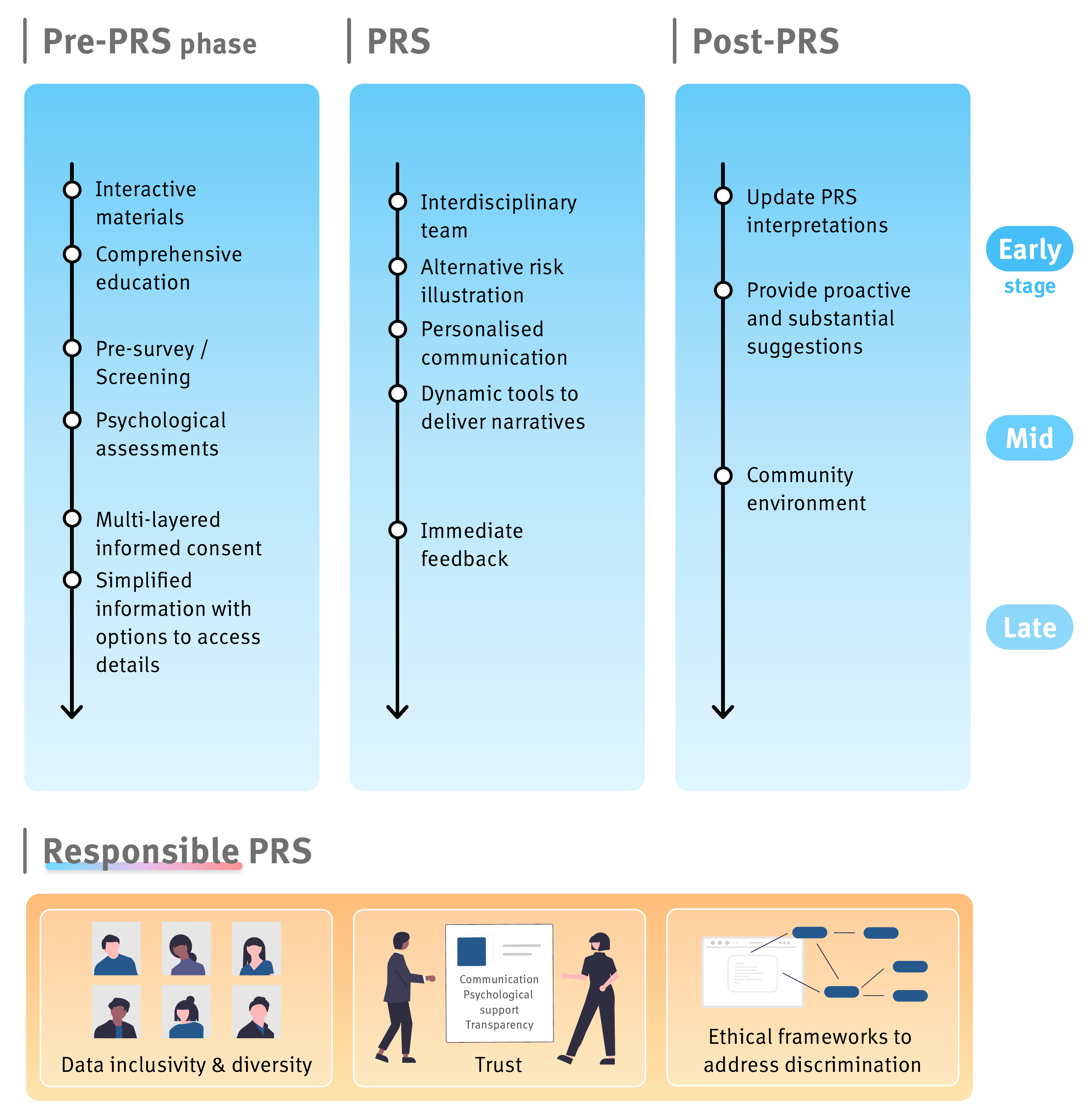}
  \caption{\textbf{Design Implications for Pre-PRS, PRS, Post-PRS Phases, and Responsible PRS.} Each phase is divided into early, mid, and late stages, offering specific implications as flexible reference points rather than a fixed or exhaustive framework.}
  \Description{The figure presents the design implications for PRS services across three phases: Pre-PRS, PRS, and Post-PRS. For the Pre-PRS phase, the design implications include interactive materials, comprehensive education, pre-survey/screening, psychological assessments, multi-layered informed consent, and simplified information with options for details. During the PRS phase, implications include an interdisciplinary team, alternative risk illustration, personalised communication, dynamic tools for narrative delivery, and immediate feedback. In the Post-PRS phase, implications involve updating PRS interpretations, providing proactive and substantial suggestions, and fostering a community environment. Additionally, responsible PRS includes efforts in data inclusivity and diversity, trust-building, and ethical frameworks to address discrimination.}
  \label{fig:design-implications}
\end{figure*}

\subsubsection{Implications} Our design implications for PRS are directly grounded in findings from our study, which includes a comprehensive review of user needs across different phases of their journey with PRS, as shown in Figure \ref{fig:design-implications}. The resulting framework, covering the \textit{Pre-PRS}, \textit{PRS}, and \textit{Post-PRS} phases, ultimately aims to the \textit{Responsible PRS}.

\textbf{Pre-PRS Phase}. During the Pre-PRS phase, users require comprehensive education about PRS, including its benefits, limitations, and processes. This need for education aligns with research on the challenges users face in understanding PRS and complex genetic information \cite{dougherty2014critical,polygenic2021responsible,lewis2022patient,torkamani2018personal}. For example, educational resources such as interactive modules, videos, or FAQs can facilitate this process, with the PRS educational web resource from the Broad Institute\footnote{\href{https://polygenicscores.org/explained/}{https://polygenicscores.org/explained/}} providing a practical example. To extend the literature on personalised health education \cite{lewis2022patient,slunecka2021implementation}, we propose pre-screening users to assess health background and psychological status, enabling tailored interventions such as one-on-one counselling. Meanwhile, implementing a pre-screening process also raises concerns about inadvertently introducing barriers to PRS access, such as for individuals with lower health literacy or limited resources. Strict or overly complex screening criteria could discourage or delay those who may benefit most \cite{balogh2015overview}. To mitigate this risk, we recommend making the screening process flexible and supportive rather than exclusionary. For instance, users identified as needing additional assistance could be offered supplementary educational modules or counselling sessions, rather than being denied access. Additionally, to prevent information overload during consent processes \cite{matthes2020too,baig2020m} and provide concise information to enhance user awareness \cite{ebert2021bolder}, we recommend a layered consent strategy, starting with simplified overviews and adding details as needed. We emphasise layered, adaptive information to meet individual needs, contributing to more effective consent.

\textbf{PRS Phase}. The PRS phase involves interpreting and communicating complex genetic information, a challenge noted extensively in prior studies \cite{brockman2021design,lewis2020polygenic,torkamani2018personal}. Interdisciplinary collaboration -- among genetic counsellors, medical experts, data scientists, and psychologists -- can enhance PRS interpretation accuracy. We suggest presenting PRS results in formats tailored to users' needs and preferences, considering diverse levels of expertise and psychological readiness. For example, while numerical results may appear straightforward, their interpretation can be overwhelming or anxiety-inducing for many users. Building upon the prior suggestion of the dichotomous interpretation (e.g., high vs. average risk) \cite{hao2022development}, we recommend using visual aids like risk charts, personalised summaries, and scenario-based examples to ensure genetic information is comprehensible and actionable. VR and AR tools, as shown in prior work \cite{van2022virtual,javaid2020virtual,woo2024m}, enhance the user experience by improving understanding, reducing anxiety, and fostering better interactions. Our approach situates these technologies within the PRS context to address users' emotional and informational needs. Additionally, an immediate feedback mechanism for real-time clarifications, such as live chats, can also enhance support. 

Notably, in the U.S., legislative provisions introduced under the 21st Century Cures Act final rule now require clinical laboratories to release test results directly to patients before a clinician reviews them \cite{21stCenturyCuresAct,CuresActFinalRule}. If such mandates were extended to PRS, users might receive complex, probabilistic genetic information without immediate professional guidance, potentially undermining the collaborative interpretation framework we propose. Most importantly, the impact of receiving PRS results before professional consultation may vary depending on the context of care and the nature of the condition\footnote{For example, in emergency or inpatient settings, early access to test results is often less problematic, as patients and their families are typically informed of the findings shortly after the results become available or at least before discharge \cite{genes202221st}. However, for conditions like Alzheimer's disease, the situation could be markedly different, as the implications of the results may be more emotionally charged and require careful contextualisation during discussions with healthcare professionals \cite{largent2022bringing}.}. In this context, we identify three key tensions: (1) balancing increased patient autonomy with the risk of misinterpretation, (2) ensuring timely and manageable clinician-patient dialogues, and (3) addressing equity concerns and resource allocation. To address these challenges, we suggest that PRS results' interfaces and policies should integrate mechanisms that support preliminary interpretation while users await professional guidance. Some examples include automated triage and educational modules, which can provide immediate insights tailored to user profiles, contextualised test results that place genetic information in an understandable framework \cite{hahne2022communication}, and on-demand support for clarifications -- allowing users to seek help until a healthcare professional can offer a more in-depth consultation. Furthermore, these initiatives might engage eight stakeholder groups that could be significantly affected by these legislative changes, as highlighted by Arvisais-Anhalt et al. \cite{arvisais2023laboratory} -- to ensure that diverse and relevant perspectives are included in the process.

\textbf{Post-PRS Phase}. Post-PRS, the dynamic nature of PRS results -- affected by ongoing genetic research, algorithmic updates, and new data -- requires users to be continuously informed \cite{torkamani2018personal,lewis2020polygenic}. Although an individual's genotype remains unchanged, evolving PRS interpretations require clear, context-specific updates\footnote{New genetic research may expand reference databases, altering PRS values without changing risk categories. Conversely, stable PRS values might be reinterpreted due to updated guidelines or medical insights, affecting recommendations.}. To address this and extend the literature on dynamic health interventions \cite{torkamani2018personal,murdoch2023mobilising}, we recommend personalised health management plans that adapt to scientific advancements, including lifestyle recommendations, preventive measures, and medical screenings that adapt as new insights emerge. Eventually, fostering supportive communities such as online forums or local support groups can provide ongoing emotional support, reduce social isolation, and serve as a feedback loop for PRS providers.

\subsubsection{Responsibility} A responsible PRS framework demands commitment, sustainability, interdisciplinary support, and significant investment, yet achieving these is inherently complex. While principles such as data inclusivity, clear communication, psychological support, transparency, and ethical use are foundational, their implementation might also face barriers. For instance, including underrepresented groups in large-scale biomedical databases, such as the UK Biobank \cite{bycroft2018uk} and the All of Us program \cite{all2019all}, is essential for enhancing data diversity and ensuring representative PRS models. Yet, the inclusion of diverse populations is resource-intensive and often hindered by geopolitical \cite{shih2023origins}, infrastructural \cite{faure2021mapping}, and ethical challenges \cite{baynam2024increasing}. Effective communication strategies are equally critical but require simplifying complex genetic information without oversimplifying or misleading, which is a difficult balance to achieve. Similarly, personalised genetic counselling is crucial for providing emotional support and helping users make informed decisions. However, scaling such services to meet population-level needs would demand significant investment in training and resources -- an unlikely scenario in regions with limited healthcare budgets. Building trust through transparency and ethical frameworks is another pillar, but trust is fragile and can be undermined by systemic inequities \cite{webster2022social} or high-profile data breaches \cite{nemec2024data}, particularly in sensitive areas like insurance or reproduction as our findings illustrated.

While a responsible PRS framework is ambitious, it offers opportunities for meaningful progress. Targeted efforts prioritising the most impactful areas -- such as improving data diversity and ethical governance -- could lay the groundwork for broader, sustainable progress in PRS development.

\subsubsection{Stakeholders} We acknowledge that the development, implementation, and maintenance of PRS require a multidisciplinary approach involving multiple stakeholders from different disciplines. Although our findings suggest that an interdisciplinary team is crucial for effectively addressing the complexities of PRS, the current understanding of these stakeholders' specific roles and contributions remains limited. Additionally, stakeholder composition may vary by PRS provider types, such as direct-to-consumer, healthcare systems, research institutions, or community initiatives. 

However, given the diversity of stakeholders involved, significant interdisciplinary challenges and tensions may arise throughout the PRS lifecycle. PRS stakeholders might have \textbf{different priorities}, such as HCI researchers emphasise user engagement and iterative design, while healthcare professionals focus on clinical accuracy and evidence-based practice \cite{blandford2018seven}. These differing perspectives challenge balancing user-centred design with clinical standards. Additionally, healthcare professionals often face \textbf{time constraints} \cite{gagnon2012systematic}, making comprehensive PRS interpretation challenging without \textit{adequate} support. Another challenge is \textbf{balancing data transparency with privacy regulations}, as users, providers, and regulators might have conflicting expectations about data control and privacy. This divergence can lead to a disconnect between user expectations for transparency and the practicalities faced by healthcare systems -- the ability to choose and control personal data is essential for the public to determine whom to trust and how that trust is managed \cite{rivas2021citizens}. Adding to these challenges is \textbf{the timing and approach to public engagement}, which involves multiple stakeholders. Engaging the public early in the development of emerging health technologies like PRS can greatly enhance the technical design process, ensuring that these innovations are more aligned with public needs and values \cite{o2012public}. Conversely, the delayed involvement can create a disconnect between technological capabilities developed by such as engineers and data scientists, and the public understanding that healthcare professionals, HCI researchers, and communicators need to foster. Consequently, it may lead to misunderstandings and unforeseen issues, requiring additional resources for education and revisions \cite{slattery2020research}. 

Future work could explore ways to enhance stakeholder integration throughout the PRS lifecycle, focusing on addressing the root causes of interdisciplinary challenges collaboratively. Aligning HCI's user-centred priorities with the clinical rigour of healthcare professionals may benefit from shared frameworks that foster deeper understanding rather than surface-level compromises \cite{blandford2018seven,agapie2024conducting}. Support systems for healthcare professionals might also be re-examined, such as improving training or developing team-based care models \cite{will2019team}. Similarly, balancing transparency with privacy might involve adaptive governance that aligns with societal values, incorporating diverse perspectives from ethicists, policymakers, and users. Eventually, public engagement could shift towards more proactive -- continuous involvement from the beginning --  potentially supported by citizen advisory boards \cite{chan2022public} or other institutional structures.

\section{Towards Human-Precision Medicine Interaction (HPMI)}

As we integrate PRS and other PM technologies into various aspects of healthcare and everyday life, it is crucial to understand not only their capabilities but also how they are perceived and used by individuals. Given that most current PM technologies are based on biological data \cite{delpierre2023precision}, technologies like PRS represent specific applications within PM that share common challenges related to the interpretation and communication of complex health data. This ongoing process involves a complex interplay between humans and PM systems, which we refer to as \textbf{Human-Precision Medicine Interaction (HPMI)}. We introduce the term HPMI here as a means to highlight interactions between individuals and PM technologies, viewed through a HCI lens. We coin this term HPMI but do not seek to fully define it here, instead ending our paper with an invitation to the HCI community to further develop HPMI as a focused area of research and practice in HCI in the future. Here, we outline two foundational considerations -- \textbf{complex health data communication and interpretability}, and \textbf{systemic collaboration and redesign} -- to initiate this area of work and guide future exploration.

One of the key goals of PM technologies is prevention, aiming to identify risks early and support proactive health measures \cite{denny2021precision}. By examining the predictive context of PRS as an example of genetic-data-based PM technologies, our research aims to understand how individuals interact with these probabilistic PM technologies and make informed decisions based on such data. Our participants expressed concerns about understanding genetic data, especially in probabilistic and predictive contexts like PRS. These challenges are indicative of broader issues faced when interacting with various PM technologies, involving making complex health data understandable for users of different expertise levels, both technical and non-technical. Therefore, we hope that HPMI addresses these challenges by rethinking how PM data is communicated to ensure complex information is both accessible and meaningful.

For the CHI and broader HCI communities, HPMI presents an opportunity to expand research by addressing the complexities of PM. Beyond predictive-nature PM technologies like PRS, HCI researchers might discover potential roles in other PM applications \cite{denny2021precision} such as personalised treatments \cite{robson2017olaparib} and patient-specific therapeutic interventions \cite{malone2020molecular}. Future work could focus on making genomic and other health data more understandable, developing interaction models that support user autonomy, addressing ethical considerations, and facilitating collaboration with stakeholders to manage tensions and conflicting interests. This requires rethinking PM technology design. Beyond enhancing usability, it involves fundamentally reconsidering how personal health data is presented, emphasising interpretability, ethical transparency, and user empowerment. As PM technologies become more accessible, the role of the CHI and HCI communities in promoting user autonomy, informed consent, and psychological well-being will be crucial in shaping the future of PM and wider healthcare interactions.

\section{Limitations and Conclusion}
We acknowledge several limitations in our study. Although focusing on a UK-based population provided valuable context, it may limit the generalisability of the findings to other regions. For example, the Global South. The small sample size of eleven interviews, although providing in-depth insights, might not capture the full diversity of perspectives. Additionally, because participants had higher education levels, they might possess different awareness, understanding, or interest in PRS services compared to the general population, potentially skewing our findings. We also recognise that the inherent complexity of PRS may have influenced participant selection, potentially deterring individuals with less familiarity or interest in such technologies, thereby introducing selection bias. Given the rapid advancements in PRS and PM, some insights may quickly become outdated, underscoring the need for continuous research to maintain relevance.

In conclusion, our study provides valuable initial insights into public perceptions and attitudes toward PRS in the UK. While participants expressed cautious optimism about the potential benefits of PRS, significant barriers remain, including concerns about data inclusivity, psychological impacts, and trust. To address these challenges, it is essential to improve education, establish robust ethical frameworks, and adopt inclusive research practices. Future research should explore public attitudes toward PRS and other PM technologies through longitudinal studies, tracking how perceptions evolve. Investigating the specific concerns of diverse demographic groups is crucial to ensure that personalised healthcare benefits everyone. Based on our findings, we proposed design implications for a responsible PRS framework and introduced the evolving concept of Human-Precision Medicine Interaction (HPMI) to address these challenges. Through further exploration, HPMI aims to bridge the PM and HCI communities, encouraging solutions that enhance public interaction with PM technologies and foster better decision-making and health outcomes.

\begin{acks}

We thank all participants for their views, insights, and contributions to this project. We extend our gratitude to Colin Farquharson for acting as an independent contact throughout the research process. We also thank Mulya Agung, Elly Gaunt, and Jing Qi Chong for their support. We appreciate the iterative feedback from our anonymous reviewers. Finally, the first author would like to acknowledge the Henry Lester Trust and the Genetics Society for the grants they awarded to support his studies.

\end{acks}

\bibliographystyle{ACM-Reference-Format}
\input{main.bbl}

\newpage
\appendix

\section{Interview Materials}

\subsection{Summaries to Ten Stories with ContraVision}
\label{sum-10story}

\aptLtoX{\begin{framed}
\textbf{Legends:} 
\begin{itemize}
\item $\rightrightarrows$ \textit{ContraVision point.}
\item $\circlearrowright$ \textit{Back to the ContraVision point.}
\item $\blacksquare$ \textit{End of the whole story.}
\end{itemize}
\end{framed}}{\fbox{%
\textbf{Legends:} 
\begin{Bitemize}
\item $\rightrightarrows$ \textit{ContraVision point.}
\item $\circlearrowright$ \textit{Back to the ContraVision point.}
\item $\blacksquare$ \textit{End of the whole story.}
\end{Bitemize}}}

\subsubsection{$\oslash1$ Privacy}
James (he/him) expressed concerns to his GP about his family's history of Type 2 diabetes. The GP suggested a PRS analysis to assess his risk, which intrigued James, though the GP couldn't provide more information as it's not officially used by the NHS yet. $\rightrightarrows$ \\ \textbf{Scenario 1} $\rightarrow$ On a PRS service website, James noticed a statement about accessing his genetic data via a secure Application Programming Interface (API). After researching APIs and learning they protect his data's intricacies, he decided to use the site to generate his PRS report. $\circlearrowright$ \\\textbf{Scenario 2} $\rightarrow$ Despite the simple user interface, James was concerned about the trustworthiness of this service, particularly regarding data privacy. Worried about potential data leaks and retention, he still chose to generate the PRS report, prioritising his health but remaining anxious about the security of his genetic information. $\blacksquare$

\subsubsection{$\oslash2$ Mental Health}Lucy (she/her) received her PRS results, which indicated a higher chance of developing asthma compared to 84\% of people in the database. She had been worrying about this for several days. $\rightrightarrows$  \\ \textbf{Scenario 1} $\rightarrow$ Lucy researched her PRS results and learned that genetics isn't the only factor in health. After consulting a friend in genetics, she decided to use her PRS as lifestyle guidance, making changes like quitting her hairstylist job to reduce asthma risk. $\circlearrowright$\\\textbf{Scenario 2} $\rightarrow$ Lucy, prone to overthinking, became obsessed with her PRS results, leading to sleepless nights and anxiety. Three months later, she was diagnosed with mild depression and regretted learning her PRS results, feeling they caused more harm than good. $\blacksquare$

\subsubsection{$\oslash3$ Interpretation}Mr Morrison (he/him) and Mrs Morrison (she/her) sought genetic counselling through the NHS as they planned to have a child, despite no family history of genetic illnesses. They began testing with their GP, who suggested additional PRS reports, though Mrs Morrison found the concept difficult to understand. $\rightrightarrows$  \\ \textbf{Scenario 1} $\rightarrow$ The GP, unable to answer Mrs Morrison's technical questions, referred her to a PRS specialist who clarified everything. They also learned that Mr Morrison's mixed ethnicity might impact his PRS results due to the database's limited diversity. Feeling informed, they decided to proceed with PRS testing. $\circlearrowright$\\\textbf{Scenario 2} $\rightarrow$ The GP couldn't answer Mrs Morrison's questions or find someone who could, leaving her to research on her own. Additionally, the GP was unsure about the impact of Mr Morrison's mixed ethnicity on PRS results. Lacking confidence in the GP's expertise, Mr and Mrs Morrison were unconvinced that the healthcare team could competently handle PRS testing. $\blacksquare$

\subsubsection{$\oslash4$ Insurance}``BritHealth'' is a healthcare insurance company in the UK. They would like to use customers' PRS results to further design their insurance products. $\rightrightarrows$  \\ \textbf{Scenario 1} $\rightarrow$ ``BritHealth'' chose to charge all customers the same insurance price, regardless of their PRS results. However, they offered extra services tailored to the customer's risk level. Customers with decreased risks received four private GP sessions annually, while those with increased risks were provided with personalised health plans, developed in collaboration with healthcare professionals. $\circlearrowright$\\\textbf{Scenario 2} $\rightarrow$ ``BritHealth'' opted to charge different insurance prices based on customers' PRS results, implementing a five-level pricing system. Customers with higher disease risks paid more, reflecting the increased likelihood of future claims, while those with lower risks paid less, with the standard price set for average risk levels. $\blacksquare$

\subsubsection{$\oslash5$ Ageing}Kevin (he/him), an 82-year-old man with a history of multimorbidity, was introduced to the PRS service during his quarterly GP visit. This was the first time he heard about PRS. Intrigued by this new service, Kevin decided to give it a try to gain insights into his genetic predispositions to certain diseases. $\rightrightarrows$  \\ \textbf{Scenario 1} $\rightarrow$ Kevin revisited his GP and received PRS reports showing a genetic predisposition for conditions he already had, offering him some relief from his previous lifestyle regrets. His partner Alison (she/her) also tried the PRS service, discovering her own genetic risks but remained sceptical about its accuracy, joking about its relevance to her current health. $\circlearrowright$\\\textbf{Scenario 2} $\rightarrow$ Kevin returned to his GP only to find that his PRS reflected the diseases he already had. He, thus, was frustrated. He expressed doubts about the service's suitability and relevance for older individuals, suggesting it might not be worth recommending if it doesn't provide actionable or beneficial insights. $\blacksquare$

\subsubsection{$\oslash6$ Ethics}Andy (he/him) and Chloe (she/her) were considering starting a family again and exploring genetic counselling to ensure their future children's health. They learned about the PRS service, which can assess the risk of passing on genetic traits that might lead to diseases like cystic fibrosis, sickle cell anaemia, and haemophilia. $\rightrightarrows$  \\ \textbf{Scenario 1} $\rightarrow$ Initially unsure about the ethics of using PRS for reproductive decisions, Andy and Chloe decided to proceed with the analysis. The results indicated a lower risk of passing on common diseases. Relieved and reassured, they went ahead and had a healthy baby girl, viewing PRS as a tool that provided mental comfort without altering their intent to have a child regardless of the results. $\circlearrowright$\\\textbf{Scenario 2} $\rightarrow$ Andy and Chloe felt that using PRS for reproductive decisions bordered on eugenics, raising ethical concerns. They worried about the implications of using genetic information to make life decisions and the potential for it to influence societal norms and access to services, viewing it as a form of modern eugenics and questioning the ethicality of applying such technology in their situation. $\blacksquare$

\subsubsection{$\oslash7$ School}Mr Smith (he/him), a school principal, was concerned about certain students who struggled academically or exhibited behavioural issues such as hostility, hyperactivity, and inattention. Interested in improving student support, he learned about PRS and used them to identify students at risk for academic and behavioural difficulties. $\rightrightarrows$  \\ \textbf{Scenario 1} $\rightarrow$ Mr Smith implemented PRS results to create personalised educational and intervention programs. Students identified with a high PRS for ADHD received tailored classroom adjustments, while those with high scores for anxiety or depression were offered counselling and peer support. He continuously monitored progress and adjusted strategies, planning to evaluate the program's effectiveness by comparing participants with non-participants. $\circlearrowright$\\\textbf{Scenario 2} $\rightarrow$ Using the PRS results, Mr Smith initiated a weekly workshop for students with intellectual disabilities. Initially, the workshops seemed beneficial, boosting confidence and social interaction among participants. However, it led to unintended negative consequences, including bullying and ridicule from peers. Concerned parents voiced their issues, prompting Mr Smith to seek alternative methods to mitigate these adverse effects. $\blacksquare$

\subsubsection{$\oslash8$ ``Next Steps''}Olivia (she/her), experiencing discomfort around her heart, consulted her GP who found no current diseases but suggested a PRS analysis to assess her genetic risk for cardiovascular disease (CVD). The results indicated Olivia had a higher likelihood of developing CVD compared to 88\% of people. $\rightrightarrows$  \\ \textbf{Scenario 1} $\rightarrow$ The GP prescribed Olivia three medications, asserting a 99\% success rate in preventing CVD if taken regularly and on time. Relieved by the proactive approach, Olivia accepted the additional daily task of medication management, feeling it was a small price to pay for potentially avoiding CVD. $\circlearrowright$\\\textbf{Scenario 2} $\rightarrow$ Without preventive medications available for CVD, the GP recommended lifestyle adjustments to Olivia, such as reducing intake of oily and fried foods, increasing dietary fibre, and maintaining physical activity. Frustrated by the ambiguity of how much lifestyle change was necessary to prevent CVD, Olivia committed to following the GP's advice but remained anxious about the sufficiency of these measures. $\blacksquare$

\subsubsection{$\oslash9$ Transparency \& Visualisation}Emily (they/them), intrigued by the PRS service after reading about it, expressed concerns about the safety of uploading sensitive genotype data online. They understood that genotype data is highly personal, likening it to a ``passcode'' for creating another version of oneself. $\rightrightarrows$  \\ \textbf{Scenario 1} $\rightarrow$ Emily discovered a new PRS service that operates locally on the user's device without requiring data upload to the internet. This service uses downloadable software that analyses genotype data directly on the user's computer, visualising the process and making it interactive and educational. Emily found the software not only fun but also informative, as it helped them understand the PRS methodology and their personal risk for developing diseases like Type 2 diabetes. $\circlearrowright$\\\textbf{Scenario 2} $\rightarrow$ Emily, seeking a trustworthy PRS service, eventually chose one that promised not to misuse or retain their data post-analysis. After uploading their data, they received a PRS report indicating their risk for five common diseases, presented in a ``normal distribution'' curve. Confused by the statistical representation and the implications of their results, Emily decided to consult a GP or genetic counsellor for a clearer understanding. $\blacksquare$

\subsubsection{$\oslash10$ Policy \& Governance}The local government is considering legislation to regulate the use of PRS services, which involve handling sensitive genotype data. The proposed regulations aim to manage how personal genotype data is used within commercial contexts and by PRS service operators. $\rightrightarrows$  \\ \textbf{Scenario 1} $\rightarrow$ The government plans to create a ``PRS Harbour,'' a centralised platform where all PRS activities are conducted. Stakeholders, including PRS service operators and third parties using PRS commercially, must operate within this hub, accessing encrypted genotype data without direct handling. An interdisciplinary team will oversee and ensure the security and ethical use of the platform. $\circlearrowright$\\\textbf{Scenario 2} $\rightarrow$ The government considers partnering with the NHS to manage PRS use, requiring ethical approval for PRS operators and commercial users, which could restrict access for smaller entities. Individuals must receive a GP's recommendation to proceed with PRS, potentially limiting access. The NHS would form a specialised team to manage PRS operations, focusing on ethical standards and data security. $\blacksquare$

\subsection{Barriers Chosen and Priorities}
\label{app:theme-pri}

\begin{table*}
\caption{\textbf{Barriers Chosen and Their Priorities Decided by Interview Participants} (n=11). Cells that are coloured indicate that the participant chose and discussed the corresponding barrier. Darker-coloured cells denote the barriers that the participant gave the highest priority. The first row (\texttt{P1}) illustrates the colour coding and the corresponding prioritisation.}
\label{table:theme-vis}
\begin{tabular}{c|p{0.5cm}|p{0.5cm}|p{0.5cm}|p{0.5cm}|p{0.5cm}|p{0.5cm}|p{0.5cm}|p{0.5cm}|p{0.5cm}|p{0.5cm}} 
\toprule
\texttt{\textbf{PID}} $\backslash$ Barrier & \multicolumn{1}{c|}{\textbf{$\oslash$1}} & \multicolumn{1}{c|}{\textbf{$\oslash$2}} & \multicolumn{1}{c|}{\textbf{$\oslash$3}} & \multicolumn{1}{c|}{\textbf{$\oslash$4}} & \multicolumn{1}{c|}{\textbf{$\oslash$5}} & \multicolumn{1}{c|}{\textbf{$\oslash$6}} & \multicolumn{1}{c|}{\textbf{$\oslash$7}} & \multicolumn{1}{c|}{\textbf{$\oslash$8}} & \multicolumn{1}{c|}{\textbf{$\oslash$9}} & \multicolumn{1}{c}{\textbf{$\oslash$10}} \\ \midrule
\texttt{\textbf{P1}} &  & \cellcolor[HTML]{4BA5ED}{2nd} & \cellcolor[HTML]{8ACFEC}{3rd} &  & \cellcolor[HTML]{D3F4FF}{4th} &  &  & \cellcolor[HTML]{1F4A6C}{\textcolor{white}{1st}} &  &  \\ \hline
\texttt{\textbf{P2}} &  &  & \cellcolor[HTML]{4BA5ED} &  &  & \cellcolor[HTML]{1F4A6C} & \cellcolor[HTML]{8ACFEC} &  &  &  \\ \hline
\texttt{\textbf{P3}} &  & \cellcolor[HTML]{4BA5ED} &  &  &  & \cellcolor[HTML]{8ACFEC} &  & \cellcolor[HTML]{1F4A6C} &  &  \\ \hline
\texttt{\textbf{P4}} &  &  & \cellcolor[HTML]{4BA5ED} &  & \cellcolor[HTML]{1F4A6C} &  &  & \cellcolor[HTML]{8ACFEC} &  &  \\ \hline
\texttt{\textbf{P5}} &  & \cellcolor[HTML]{D3F4FF} &  & \cellcolor[HTML]{8ACFEC} & \cellcolor[HTML]{1F4A6C} &  &  &  &  & \cellcolor[HTML]{4BA5ED} \\ \hline
\texttt{\textbf{P6}} &  & \cellcolor[HTML]{4BA5ED} &  &  & \cellcolor[HTML]{8ACFEC} &  &  & \cellcolor[HTML]{1F4A6C} &  &  \\ \hline
\texttt{\textbf{P7}} & \cellcolor[HTML]{1F4A6C} & \cellcolor[HTML]{4BA5ED} &  & \cellcolor[HTML]{D3F4FF} & \cellcolor[HTML]{D3F4FF} &  &  &  &  & \cellcolor[HTML]{8ACFEC} \\ \hline
\texttt{\textbf{P8}} &  & \cellcolor[HTML]{4BA5ED} &  &  &  & \cellcolor[HTML]{8ACFEC} &  &  &  & \cellcolor[HTML]{1F4A6C} \\ \hline
\texttt{\textbf{P9}} & \cellcolor[HTML]{1F4A6C} & \cellcolor[HTML]{4BA5ED} &  &  &  &  & \cellcolor[HTML]{8ACFEC} &  &  &  \\ \hline
\texttt{\textbf{P10}} &  & \cellcolor[HTML]{4BA5ED} &  &  & \cellcolor[HTML]{8ACFEC} &  &  & \cellcolor[HTML]{1F4A6C} &  &  \\ \hline
\texttt{\textbf{P11}} &  &  &  & \cellcolor[HTML]{1F4A6C} & \cellcolor[HTML]{8ACFEC} &  &  &  & \cellcolor[HTML]{4BA5ED} &  \\ \bottomrule 
$\oslash$ &\centering\rotatebox[origin=c]{270}{Data privacy} & \centering\rotatebox[origin=c]{270}{Mental health} & \centering\rotatebox[origin=c]{270}{Interpretation} & \centering\rotatebox[origin=c]{270}{Insurance} & \centering\rotatebox[origin=c]{270}{Ageing} & \centering\rotatebox[origin=c]{270}{Ethics} & \centering\rotatebox[origin=c]{270}{School} & \centering\rotatebox[origin=c]{270}{``Next steps''} & \centering\rotatebox[origin=c]{270}{Transparency} & \rotatebox[origin=c]{270}{Policy}   \\ \bottomrule

\end{tabular}

\end{table*}

\end{CJK}
\end{document}

%% file: main.bbl
%%% -*-BibTeX-*-
%%% Do NOT edit. File created by BibTeX with style
%%% ACM-Reference-Format-Journals [18-Jan-2012].